\documentclass[a4paper,twocolumn,11pt,unpublished=2017-05-09]{quantumarticle}
\pdfoutput=1
\usepackage[utf8]{inputenc}
\usepackage[english]{babel}
\usepackage[T1]{fontenc}
\usepackage{nicefrac}
\usepackage{amsmath}
\usepackage{amsfonts}
\usepackage{amssymb}
\usepackage{physics}
\usepackage{hyperref}
\usepackage{array}
\usepackage{float}
\usepackage{cite}

\usepackage{tikz}
\usepackage{lipsum}

\begin{document}

\title{Towards the coherent control of robust spin qubits in quantum algorithms}

\author{Luis Escalera-Moreno}
\affiliation{Max Planck Institut für Quantenoptik, Garching bei München, 85748, Germany}
\orcid{0000-0001-8015-8606}
\email{luis.escalera.moreno@mpq.mpg.de}
\maketitle

\begin{abstract}
  Many efforts have succeeded over the last decade at lengthening the timescale in which spin qubits loss quantum information under free evolution. With these design principles at a mature stage, it is now timely to widen the scope and take the whole picture: concerning applications that require user-driven coherent evolutions, qubits should be assessed operating within the desired algorithm. This means to test qubits under external control while relaxation and imperfections are active, and to maximize the algorithm fidelity as the actual figure of merit. Herein, we pose and analytically solve a master equation devised to run one-spin-qubit gate-based algorithms subject to relaxation. It is handled via a home-made code, QBithm, which inputs gate sequences and relaxation rates thus connecting with the longstanding work devoted to their \textit{ab initio} computation. We evaluate the impact of relaxation and potential experimental imperfections in the calculated fidelities, and implement well-known pulse sequences quantitatively agreeing with experimental data. Hopefully, this work will stimulate the study of many-spin-qubit systems in quantum algorithms, and will serve as a help to design robust spin qubits against decoherence and to perform better-characterized experiments.
\end{abstract}

\section{Introduction}

Quantum simulation and quantum computation are expected to circumvent the exponential scaling that makes problems of wide interest -but with a large input- be unsolvable on classical computers. Instances of these problems include the prediction of materials relevant for industry and society, and the resolution of both optimization and combinatorics queries for policy-making. While this picture seems promising at establishing new scientific and technological revolutions, the required experimental techniques rely on noisy operations that eventually spoil quantum information. A clear path towards a prototype able to solve all the above-mentioned problems in the mid term is nowadays elusive. Until quantum error correction protocols can be implemented in the lab on a massive scale, the provisory solution is that of operating on noisy intermediate-scale quantum (NISQ) devices. These are the platforms currently devoted to test algorithms with quantum advantage,\cite{Intro20202} namely the ones which, by exploiting quantum properties, offer a more efficient scaling for those specific tasks where any conceivable classical algorithm would never do any better than exponentially.\cite{Intro20214}
\medbreak
The logical operations or gates that compose a quantum algorithm are mapped onto manipulations of a given physical degree of freedom: the ground energy spectrum of an atom or an ion, the charge or the flux of a carrier, the polarization of a photon, or the motion of a mechanical oscillator.\cite{Intro20201,Intro20171,Intro2019,Intro20211} In particular, beside technological applications such as sensing and communication,\cite{Review2021,Review2022} the spin is one of the best-suited candidates for quantum computation,\cite{Intro20212} where the minimum amount of information -termed as qubit- is physically realized in the spin states of defect centers,\cite{Intro20213,Intro20172,Intro2018} donor atoms in silicon,\cite{Intro2022}, and magnetic molecules.\cite{Review2019,Review20201,Review20202,Imaging20211,Imaging20213} The proved potential of this degree of freedom in terms of initialization, long-lived coherent control, readout, single-spin manipulation, scalability and applications can be further enhanced within the molecular approach, where the well-developed tools of synthetic chemistry are ideal to tune and improve \textit{at will} the properties of the so-called molecular spin qubits embedded in magnetic molecules.\cite{Qudits20211,Qudits20213,Optimal20221,Optimal20222,Puertas20222,Imaging20172,Puertas2017,Escalado2021,QEC20222,Overview2024}
\medbreak
An important conundrum faced over the last decade from both theory and experiment has been to unveil those mechanisms that promote relaxation in a spin under free evolution.\cite{Overview2020,Overview2021,Overview2022} This has been crucial to deepen understanding of the phase memory time $T_m$, i.e. the timescale for the survival of quantum information stored in a freely-evolving spin qubit. While the devoted efforts have yielded key design principles to lengthen $T_m$, one must not forget that this figure of merit assesses spin qubits under free evolution -as a quantum memory- thus mostly neglecting their performance in the important family of applications where they are coherently user-driven. Since spin qubits are also envisaged for being employed in quantum algorithms, the focus should eventually be put on user-driven evolutions and on how spin qubits perform within the algorithm of interest. These remarks should promote the algorithm fidelity as the actual figure of merit which tests not only relaxation but also other sources of error -such as experimental imperfections- whose impact has gone unheeded so far and makes $T_m$ be insufficient when characterizing the qubit performance. Hence, in the spirit of benchmarking the said performance, our work meets the goal of developing certification protocols for quantum algorithms and quantum devices.

\section{Summary section}

\begin{figure}[H]
    \centering
    \includegraphics[scale=0.11]{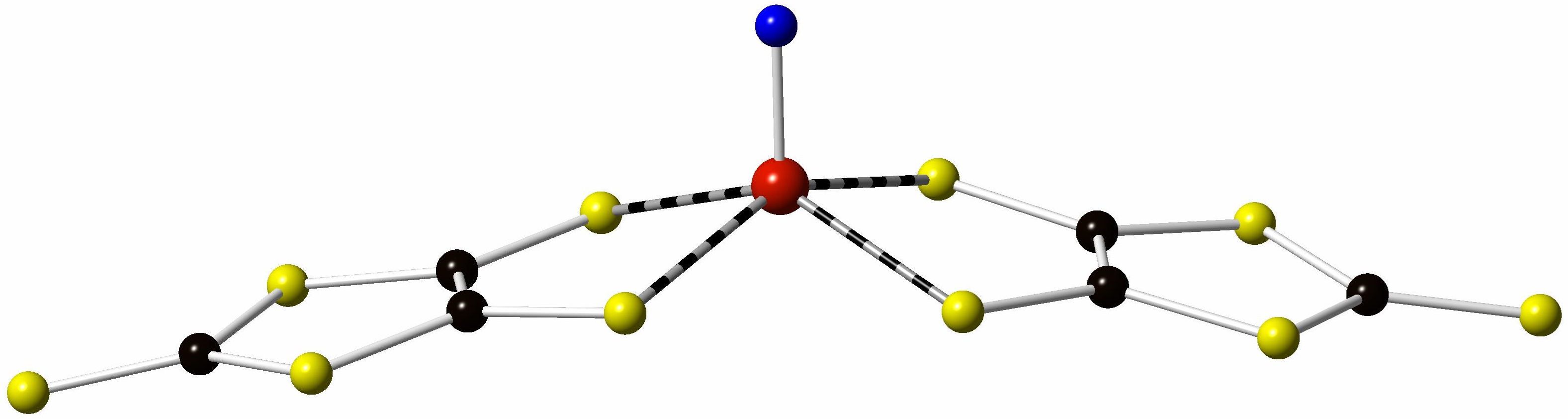}
    \includegraphics[scale=0.13]{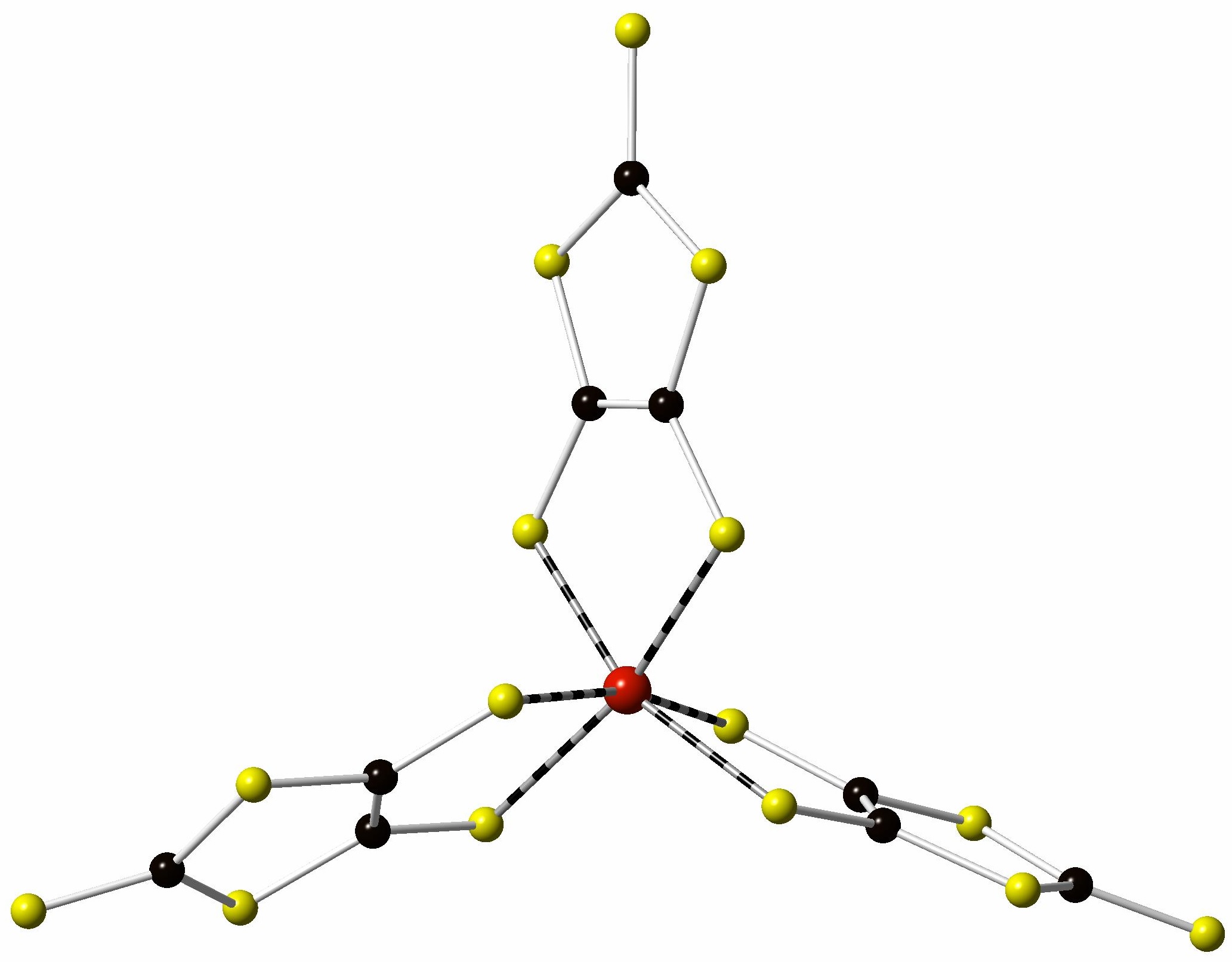}
    \includegraphics[scale=0.24]{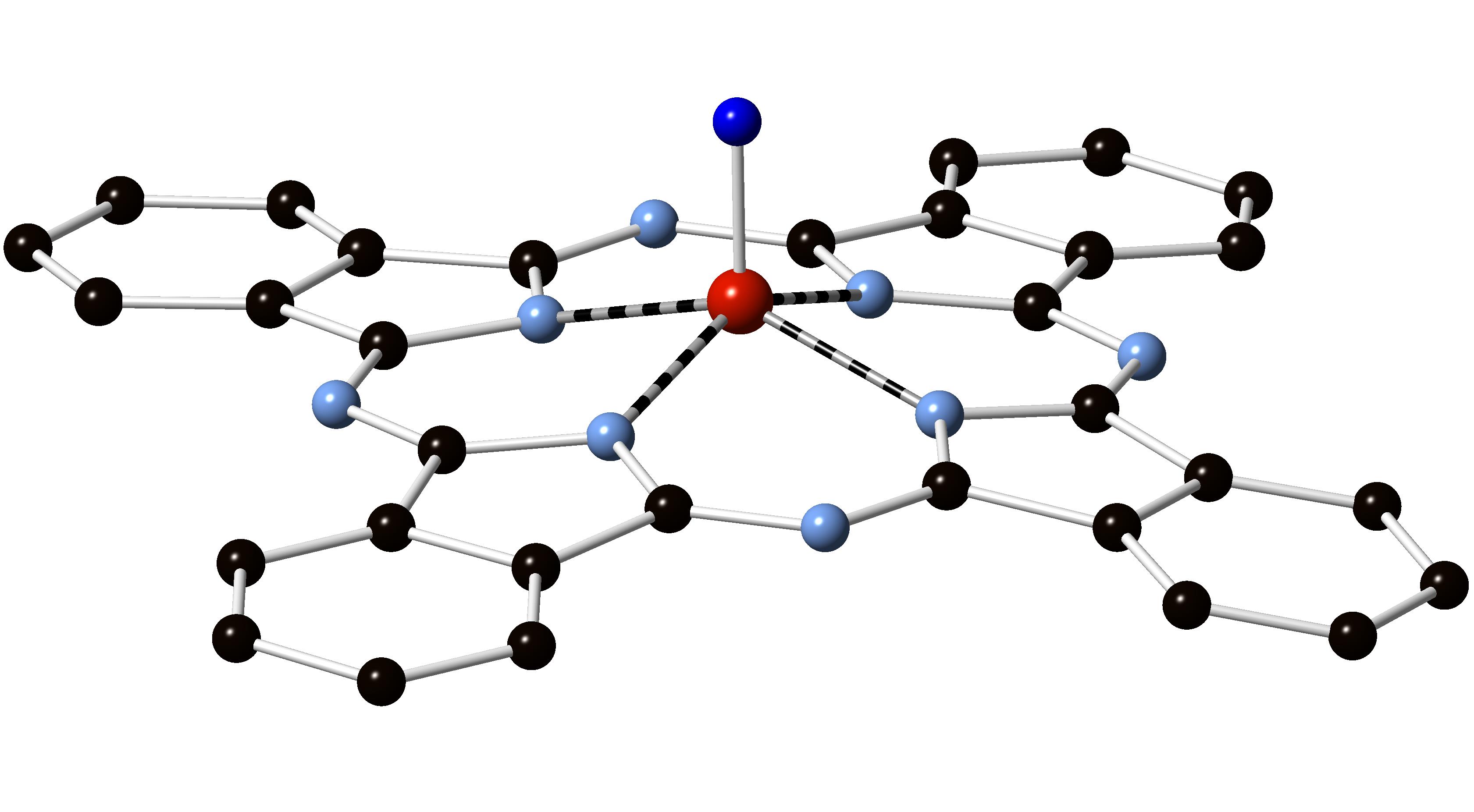}
    \includegraphics[scale=0.21]{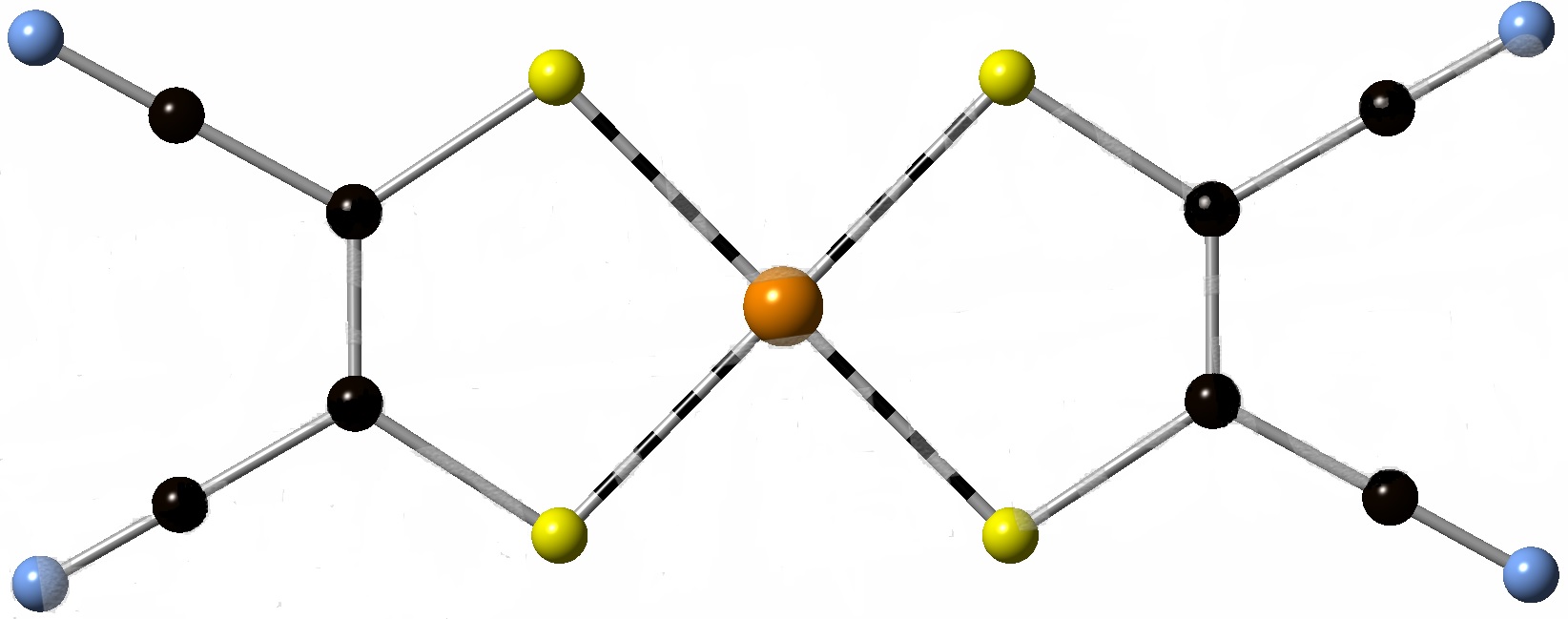}
   \caption{Molecular coordination complexes, with magnetic atomic ions V$^{4+}$ (red) and Cu$^{2+}$ (orange), as case studies of representative spin qubits. Top: \textbf{(1)}, Middle Top: \textbf{(2)}, Middle Bottom: \textbf{(3)}, Bottom: \textbf{(4)}. C: black, S: yellow, N: pale blue, O: dark blue.\cite{Sistemas20162,Sistemas20161,Sistemas2014,Sistemas2017}}
   \label{moleculas}
\end{figure}

In this manuscript, we design a master equation aimed at time-evolving the density matrix of one spin qubit, where both the main relaxation rates and the user-driven control are inputted. This allows not only studying the problem faced over the last decade, namely to track the loss of quantum information as a magnetization decay in time when the spin freely evolves under relaxation, but also to tackle what the next natural step should be: the inclusion of the said user control to drive the spin in such a way that (i) free evolutions and rotations are combined as a gate sequence to place the qubit at those desired points of the Bloch sphere thus implementing any digital one-qubit algorithm, while (ii) the main relaxation mechanisms extensively studied so far and imperfections are active. We also contribute with the analytical resolution of the master equation. This facilitates an in-depth study of the qubit dynamics unlike when one follows the more extended procedure of employing numerical methods.  
\medbreak
The relaxation rates are rather an input of our master equation and hence this work connects in a natural way with all the theoretical efforts made over the past years devoted to compute them from \textit{ab initio} methods.\cite{espesp2016,Maestra2017,espesp2019,Maestra20191,espesp20201,espesp20202,espesp20221,espesp20222,espesp2023,espvib20173,espvib2018,espvib2019,espvib20211,espvib20212,espvib20223,espvib20224,espvib2023,espvib20171,espvib20172,espvib2020,espvib20221,Maestra20194} In the race for the highly-prized scalable architectures, the realization of many-qubit systems will certainly depend on a proper understanding of the single-qubit dynamics under relaxation, user control, and imperfections as a key building block. Hence, we expect that our contribution will help to lay the foundations in the modeling of, firstly, spin-qubit pairs for logical gates and, then, quantum algorithms involving larger numbers of qubits.
\medbreak
The master equation is handled via QBithm, an open-source home-made code where one-spin-qubit algorithms are established in the input as a gate sequence (see SI). We first demonstrate control over the whole Bloch sphere, and realize one-qubit gates whose fidelity is tested against relaxation rates and experimental imperfections. Then, we run well-known pulse sequences as instances of one-spin-qubit algorithms, and obtain quantitative agreement with the experimentally-reported Rabi oscillations and spin relaxation times $T_1$, $T_m$, $T_{dd}$ (dynamical-decoupling) of case studies representing key milestones in the field of molecule-based spin qubits. Namely, those encoded in the magnetic molecules shown in Figure~\ref{moleculas}: \textbf{(1)} [VO(dmit)$_2$]$^{2-}$, \textbf{(2)} [V(dmit)$_3$]$^{2-}$, \textbf{(3)} VOPc, \textbf{(4)} [Cu(mnt)$_2$]$^{2-}$,\cite{Sistemas20162,Sistemas20161,Sistemas2014,Sistemas2017} where $J=1/2$, $I(\text{V}^{4+})=7/2$, $I(\text{Cu}^{2+})=3/2$, and the qubit base states $\ket{m_J,m_I}$ are defined from the eigenstates $\ket{-1/2,-1/2}$, $\ket{+1/2,-1/2}$ of \textbf{(1)}-\textbf{(3)}, and $\ket{-1/2,+3/2}$, $\ket{+1/2,+3/2}$ of \textbf{(4)}.

\section{Theoretical Development}

\subsection{Master Equation}

Our goal is to model the non-unitary time evolution of the spin qubit when its reduced density operator $\rho(t\geq 0)=\text{Tr}_{\text{bath}}(\rho_{\text{qubit+bath}})$ is driven through a gate sequence under the influence of the most important relaxation sources affecting spin qubits, namely the vibration and spin baths. The spin qubit is pictured as an effective doublet whose states, termed as $\ket{u_-}$ and $\ket{u_+}$ with energies $u_-<u_+$, are selected from the energy scheme of the spin Hamiltonian $H$ employed to describe the magnetic object containing the spin qubit. We model such a time evolution by making use of the GKLS equation, also known as Lindblad master equation, which determines the effective dynamics within the subspace $\{\ket{u_+},\ket{u_-}\}$ and, after adaptation to our particular problem, reads as (see SI):

\begin{align}
\frac{\partial\rho}{\partial t}=\frac{1}{i\hbar}[H_{\text{eff}},\rho]+\mathcal{L}_{\text{1-p}}\rho+\mathcal{L}_{\text{2-p}}\rho+\mathcal{L}_{\text{mag}}\rho
\label{mastereq}
\end{align}

The first term on the right-hand side of Eq.\eqref{mastereq} represents the unitary part of the qubit time evolution, while the remaining terms are the ones that induce the non-unitary dynamics on the qubit due to the interaction with the baths. $\mathcal{L}_{\text{1-p}}\rho$ and $\mathcal{L}_{\text{2-p}}\rho$ account for all the one- and two-phonon processes including both real and virtual ones. These processes altogether are characterized by temperature-dependent absorption $\Gamma_{\text{ab}}\geq 0$ and emission $\Gamma_{\text{em}}\geq 0$ rates determining the flow of spin population between $\ket{u_-}$ and $\ket{u_+}$ driven by the vibration bath.\cite{Maestra2018} $\mathcal{L}_{\text{mag}}\rho$ describes the effect of the spin bath on the qubit dynamics.\cite{Maestra2015} This bath is modeled as an isotropic magnetic noise -only dependent on the qubit-spin distances- and its effect is quantified by means of a rate $\Gamma_{\text{mag}}\geq 0$ whose magnitude is proportional to the volume concentration of spins in the bath.\cite{espesp2019} All rates are time-independent and non-negative.
\medbreak
Given a time interval $t_{i-1}<t_i$ with $1\leq i\leq n$ and $t_0=0$, $t_n$ the initial and final evolution times, the effective Hamiltonian $H_{\text{eff}}$ describes the spin qubit as a two-level system with an energy gap $u_+-u_-=\hbar\omega_{+-}$ -being $\omega_{+-}$ the Larmor angular frequency- that is either freely evolved or driven in $t_{i-1}\leq t\leq t_i$ depending on whether the Rabi frequency $\Omega_{\text{R}}=\va{\mu}_{+-}\cdot\va{B}_1/\hbar\in\mathbb{C}$ is zero or different from zero:

\begin{align}
\frac{H_{\text{eff}}}{\hbar}=\frac{\omega_{+-}}{2}\left(\ket{u_+}\bra{u_+}-\ket{u_-}\bra{u_-}\right)+
\label{Heff}
\end{align}
\begin{align*}
+\left(\Omega_{\text{R}}^*\text{cos}(\omega_{\text{MW}}\Delta t^i)\ket{u_+}\bra{u_-}+\text{c.c.}\right)
\end{align*}

The magnetic dipolar moment associated to the transition $\ket{u_-}\leftrightarrow\ket{u_+}$ is $\va{\mu}_{+-}$, $\va{B}_1$ is a linearly-polarized oscillating magnetic field with angular frequency $\omega_{\text{MW}}$, and $\Delta t^i=t-t_{i-1}$. Essentially, if the initial qubit state $\ket{\psi}$ is parameterized in terms of the zenithal $\theta$ and azimuthal $\phi$ angles of the Bloch sphere, $\ket{\psi}=\text{cos}(\theta/2)\ket{u_-}+\text{e}^{i\phi}\text{sen}(\theta/2)\ket{u_+}$ with $\ket{0}\equiv\ket{u_-}$ and $\ket{1}\equiv\ket{u_+}$, $\theta$ is controlled by activating $\va{B}_1$ for an appropriate time, while $\phi$ is changed by switching $\va{B}_1$ off and letting the Larmor precession make the qubit rotate around the positive $\mathbb{Z}^+$ axis of the said sphere with angular frequency $\omega_{+-}$. The fact that $H_{\text{eff}}$ is time-dependent hampers the analytical resolution of Eq.\eqref{mastereq}. We thus devise a clever change of picture -a unitary operator $\mathcal{U}$ transforming $\rho$ into $\overline{\rho}=\mathcal{U}\rho\mathcal{U}^{\dagger}$- to remove the said time-dependency. The resulting Hamiltonian $\overline{H}_{\text{eff}}$ -with no time-dependency- is dependent on the detuning $\delta=\omega_{+-}-\omega_{\text{MW}}$.
\medbreak
The solutions of the transformed Eq.\eqref{mastereq} can be classified in two groups depending on whether $\Gamma_{\text{ab}}=\Gamma_{\text{em}}=\Gamma_{\text{mag}}=0$ or at least one of these rates is non-zero. In the first one, no relaxation takes place and the qubit behaves as a closed system with a unitary dynamics. The solution is given by $\overline{\rho}(t_{i-1}\leq t\leq t_i)=\mathcal{R}\overline{\rho}(t_{i-1})\mathcal{R}^{\dagger}$, being $\mathcal{R}=\text{exp}\left(-i\Delta t^i\overline{H}_{\text{eff}}/\hbar\right)$ the rotation operator, or $\overline{\ket{\psi}}(t_{i-1}\leq t\leq t_i)=\mathcal{R}\overline{\ket{\psi}}(t_{i-1})$ for pure states. $\mathcal{R}$ depends on the generalized Rabi frequency $\Omega_{\text{g}}=\sqrt{|\Omega_{\text{R}}|^2+\delta^2}$ and on the rotation angle $\Omega_{\text{g}}\Delta t_i$ with $\Delta t_i=t_i-t_{i-1}$ the time taken by the rotation.

\medbreak
When $\delta=0$, $\mathcal{R}$ becomes a 2D rotation matrix and, if the qubit is described by the initial state $\ket{\psi}=\text{cos}(\theta/2)\ket{u_-}+\text{e}^{i\phi}\text{sen}(\theta/2)\ket{u_+}$, $\mathcal{R}$ rotates $\ket{\psi}$ an angle $\Omega_{\text{g}}\Delta t_i$ around the axis perpendicular to the plane containing $\ket{\psi}$, $\mathcal{R}\ket{\psi}$, and the origin $\mathcal{O}$ of the Bloch sphere. In case the rotation axis is contained in the equatorial plane of the said sphere, one finds $\mathcal{R}\ket{\psi}=\text{cos}((\theta+\Omega_{\text{g}}\Delta t_i)/2)\ket{u_-}+\text{e}^{i\phi}\text{sen}((\theta+\Omega_{\text{g}}\Delta t_i)/2)\ket{u_+}$. Namely, the rotation has been performed along the meridian given by the unaltered azimuthal angle $\phi$ and the rotation angle $\theta+\Omega_{\text{g}}\Delta t_i-\theta=\Omega_{\text{g}}\Delta t_i$ is here a zenithal angle. 
\medbreak
When $\ket{\psi}=\ket{u_-}$ ($\theta=0$), this often defines the starting point of EPR-pulse experiments in which the qubit is initialized in its lowest energy state. If $\Delta t_i$ is such that $\Omega_{\text{g}}\Delta t_i=\pi/2$ or $\Omega_{\text{g}}\Delta t_i=\pi$, $\mathcal{R}$ corresponds to the well-known $\pi/2$ and $\pi$ pulses: the first one creates an equally-weighted superposition $\ket{\text{ew}}$ between $\ket{u_-}$ and $\ket{u_+}$ contained in the equatorial plane of the Bloch sphere, while the second one transfers all the spin population from $\ket{0}\equiv\ket{u_-}$ to $\ket{1}\equiv\ket{u_+}$. If $\delta\neq0$, $\mathcal{R}(\Omega_{\text{g}}\Delta t_i=\pi/2)\ket{u_-}$ and $\mathcal{R}(\Omega_{\text{g}}\Delta t_i=\pi)\ket{u_-}$ will differ from $\ket{\text{ew}}$ and $\ket{u_+}$, resp. The realization of a rotation in such a way that the resulting state lies close enough to the expected one crucially depends on how small $\delta$ is. Hence, $\delta$ should be taken as a key imperfection to minimize.
\medbreak
The other group of solutions, also run for a finite time $\Delta t_i$, can be organized in two subgroups depending on whether $|\va{B}_1|=\omega_{\text{MW}}=0$, or $|\va{B}_1|\neq 0$ and $\omega_{\text{MW}}\neq 0$. These subgroups constitute the two building-blocks that we will employ sequentially in the form of gates for the construction of one-qubit algorithms. The first gate is known as free evolution: the qubit evolves for a given time subject to relaxation but with no user-induced driving. This evolution can be encountered (i) during the waiting time between pulses such as the $\pi/2$ and $\pi$ pulses of the Hahn sequence and (ii) after the said $\pi$ pulse of the same sequence until recording the Hahn echo.
\medbreak
The gate with $|\va{B}_1|\neq 0$, $\omega_{\text{MW}}\neq 0$ is known as rotation: the qubit is now driven by the user but, since relaxation is active, its evolution cannot be described in terms of a pure state circulating on the Bloch sphere anymore. However, if $\Delta t_i$ is short enough compared to the relaxation timescale $t_r$ set by $\Gamma_{\text{ab}}$, $\Gamma_{\text{em}}$, $\Gamma_{\text{mag}}$, the qubit evolves quasi-unitarily and the Bloch sphere picture can still be recovered. This situation applies in the description of the $\pi/2$ and $\pi$ pulses which take a few tens of nanoseconds in standard EPR experiments, while typical relaxation times in spin qubits lie in the $\mu s$ scale or above. Under this circumstance, the solution $\overline{\rho}(t=t_i)$ can be approximated by the one obtained with $\mathcal{R}$ at $t=t_i$. If additionally $\delta\approx 0$, $\mathcal{R}=\mathcal{R}(\zeta)$ is often directly expressed in terms of the rotation angle $\zeta$, with no presence of $\Omega_{\text{g}}$ and $\Delta t_i$ thus assuming that the latter two are properly chosen in the lab to produce the desired value of $\zeta$.\cite{Maestra20191}
\medbreak
Our solution $\overline{\rho}(t=t_i)$ generalizes that produced by $\mathcal{R}(\zeta)$ and provides a more realistic description of rotations since $\overline{\rho}(t=t_i)$ (i) always contains the effect of relaxation -whether small or large- after $\Delta t_i$, (ii) can deal with $\delta\neq 0$, and (iii) instead of directly using $\zeta$, full control is provided by letting the user input $|\va{B}_1|$, $\omega_{\text{MW}}$ (both determining $\delta$ and $\Omega_{\text{g}}$), and $\Delta t_i$. This allows studying deviations in the latter three as a new set of imperfections. Moreover, the fact that relaxation becomes significant when $\Delta t_i\gtrsim t_r$ precludes the use of $\mathcal{R}$ at $t=t_i$ and obliges to employ $\overline{\rho}(t=t_i)$. This is the case of Rabi oscillations where the rotation -or nutation- ranges from $\Delta t_1\sim t_0$ to $\Delta t_1\sim t_r$. While $\overline{\rho}(t_0\leq t\leq t_1)$ is able to reproduce the relaxation-induced damping of these oscillations as $t$ goes by, $\mathcal{R}$ would just produce non-damped oscillations.
\medbreak
Given the initial state $\overline{\rho}(t=t_0)=\ket{\psi}\bra{\psi}$, any digital one-qubit algorithm is constructed as a sequence $\{G_i\}_{i=1}^n$ of free evolutions and rotations: $\overline{\rho}(t=t_0)\xrightarrow{G_1,\Delta t_1}\cdots\xrightarrow{G_{i-1},\Delta t_{i-1}}\overline{\rho}(t_{i-1})\xrightarrow{G_i,\Delta t_i}\overline{\rho}(t_i)\xrightarrow{G_{i+1},\Delta t_{i+1}}\cdots\xrightarrow{G_n,\Delta t_n}\overline{\rho}(t_n)$, where each $G_i$ is activated for $\Delta t_i$ and transforms $\overline{\rho}(t_{i-1})$ into $\overline{\rho}(t_i)$ via Eq.\eqref{mastereq}. In case of not being interested in any observable, one can either (i) just follow the evolution of $\overline{\rho}$ after each $G_i$ or (ii) repeat the same algorithm with $\Gamma_{\text{ab}}=\Gamma_{\text{em}}=\Gamma_{\text{mag}}=0$ in order to obtain a new result $\overline{\rho}^0(t_n)$ and calculate the fidelity between $\overline{\rho}(t_n)$ and $\overline{\rho}^0(t_n)$ as a performance measure of the qubit in the algorithm (see \textbf{Quantum gates for one-qubit algorithms}).
\medbreak
Now, given an observable $O$, its expectation value $\langle O\rangle(t_n)$ at $t=t_n$ can be determined as $\langle O\rangle(t_n)=\text{Tr}[O\rho(t_n)]$ where $\rho(t_n)=\mathcal{U}^{\dagger}(\Delta t_n)\overline{\rho}(t_n)\mathcal{U}(\Delta t_n)$. Two observables of interest are the longitudinal $\text{M}_z=\sigma_z=\ket{u_+}\bra{u_+}-\ket{u_-}\bra{u_-}$ and in-plane $\text{M}_{xy}=\sigma_x+i\sigma_y=2\ket{u_+}\bra{u_-}$ magnetizations as they are employed in the determination of Rabi oscillations and $T_1$, $T_m$ (see \textbf{Pulse sequences: Rabi oscillations and spin relaxation times}). Note that $\langle\text{M}_{xy}\rangle(t_n)$ is complex. In order to compare with experimental data, we use $|\langle\text{M}_{xy}\rangle(t_n)|$.
\medbreak
Whenever $G_i$ is a free evolution, one can analytically find $\overline{\rho}(t_i)$ as an explicit function of $\overline{\rho}(t_{i-1})$. In particular, if $G_{i=n}$ is a free evolution, one finds the explicit relation between $\Gamma_{\text{ab}}$, $\Gamma_{\text{em}}$, $\Gamma_{\text{mag}}$, and the decay rates of $\langle\text{M}_z\rangle(t_n)$, $|\langle\text{M}_{xy}\rangle(t_n)|$ plotted vs $\Delta t_n$ (see SI). This is of interest since the last gate in the pulse sequences employed to determine the decay timescales $T_1$ and $T_m$ of $\langle\text{M}_z\rangle$ and $|\langle\text{M}_{xy}\rangle|$ consists in a free evolution with a variable time. We find that $\langle\text{M}_z\rangle(t_n)$ decays single-exponentially with a rate $\Gamma_1=\Gamma_{\text{ab}}+\Gamma_{\text{em}}+\Gamma_{\text{mag}}$, while the decay of $\langle\text{M}_{xy}\rangle(t_n)$ is double-exponentially with rates $\Gamma_1$ and $\Gamma_2=(\Gamma_{\text{ab}}+\Gamma_{\text{em}})/2+\Gamma_{\text{mag}}$.
\medbreak
We also find that the decay of the real and imaginary part of $\langle\text{M}_{xy}\rangle(t_n)$ is oscillatory with angular frequency $\omega_{+-}$. When the spin that encodes the qubit is coupled to a neighboring nuclear spin, e.g. that of $^1$H, the experimental $|\langle\text{M}_{xy}\rangle|$ may exhibit an oscillatory decay but now with the nuclear Larmor frequency of $^1$H. This situation whereby the Larmor frequency of an external spin coupled to the spin qubit determines the oscillation frequency of the mentioned magnetization decay is beyond the reach of our model. A proper description would require to expand the qubit 2D Hilbert space to include the coupled nuclear spin in the dynamics.

\subsection{Target systems}

The calculation of $\va{\mu}_{+-}$, $\Gamma_{\text{ab}}$, $\Gamma_{\text{em}}$, $\Gamma_{\text{mag}}$ requires the energy scheme -including $\ket{u_-}$, $\ket{u_+}$, $u_-$, $u_+$- of the spin Hamiltonian $H$ that models the magnetic object encoding the spin qubit (see Ref.~\citenum{espesp2016,espesp2019} and SI). We focus on the scheme produced by a ground electron spin quantum number $S\geq 1/2$. In order to include spin-orbit interactions with an orbital quantum number $L\neq 0$, we henceforth replace $S$ by the total electron spin $J$ as a good quantum number, which contains the particular case $L=0$, $J=S$. We also neglect interactions between $J$ and all possible excited quantum numbers $J_{\text{ex}}$. This means that any working temperature $T$ should be low enough compared to the energy gap between $J$ and the first $J_{\text{ex}}$. The spin Hamiltonian $H$ reads as:

\begin{align}
H=\sum_{k=2,4,6}\sum_{q=-k,...,+k}B_k^qO_k^q+
\label{spinH}
\end{align}
\begin{align*}
+\frac{\mu_B}{\hbar}\sum_{\alpha=x,y,z}g_{\alpha}B_{\alpha}J_{\alpha}+\sum_{\beta=x,y,z}A_{\beta}I_{\beta}J_{\beta}+PI_z^2
\end{align*}

The quantum number $J$ -associated to the ground electron spin operator $\va{J}=(J_x,J_y,J_z)$- is combined with three types of interaction. First, an important class of objects for us is that of molecular coordination complexes, where one or several magnetic atoms or magnetic atomic ions are surrounded by and linked to a set of donor atoms. The electrostatic field produced by the donor atoms lifts the initial degeneracy of the $2J+1$ states thus producing a zero-field splitting. This fact is accounted for with the inclusion of the Crystal Field term, where $B_k^q$ and $O_k^q$ are the Crystal Field Parameters and the Extended Stevens Operators. Second, an external, weak, and static magnetic field $\va{B}=(B_x,B_y,B_z)$ in the form of a Zeeman term, being $\mu_B$ the Bohr magneton and $g_{\alpha}$ the electron Landé factors. The third term is the hyperfine interaction between $\va{J}$ and the nuclear spin operator $\va{I}=(I_x,I_y,I_z)$ -associated to a ground nuclear spin quantum number $I$- with parameters $\{A_{\beta}\}_{\beta=x,y,z}$; while the fourth one models a zero-field splitting of $I$.
\medbreak
$H$ encompasses a large set of wide-interest systems defining spin qubits. Well-known instances are quantum dots in hetero-structures such as GaAs, vacancy centers in SiC and nanodiamonds, group VA dopant atoms in silicon, transition metal and rare-earth impurities in ionic crystals, and molecular coordination complexes with a single magnetic atomic ion like V$^{4+}$, Cu$^{2+}$, Ln$^{3+}$ (Ln = Gd, Tb, Dy, Ho). The qubit states $\ket{u_-}$ and $\ket{u_+}$ are selected among the $(2J+1)(2I+1)$ eigenstates of Eq.\eqref{spinH}.
\medbreak
In order to incorporate relaxation processes not included in $\Gamma_{\text{ab}}$, $\Gamma_{\text{em}}$, $\Gamma_{\text{mag}}$, we add $\Gamma_{\text{ab,add}}\geq 0$, $\Gamma_{\text{em,add}}\geq 0$, $\Gamma_{\text{mag,add}}\geq 0$ up to $\Gamma_{\text{ab}}$, $\Gamma_{\text{em}}$, $\Gamma_{\text{mag}}$ such that $\Gamma_{\text{a}}:=\Gamma_{\text{ab}}+\Gamma_{\text{ab,add}}$, $\Gamma_{\text{e}}:=\Gamma_{\text{em}}+\Gamma_{\text{em,add}}$, $\Gamma_{\text{m}}:=\Gamma_{\text{mag}}+\Gamma_{\text{mag,add}}$ are the actual rates employed in Eq.\eqref{mastereq} with $\Gamma_{\text{ab,add}}$, $\Gamma_{\text{em,add}}$, $\Gamma_{\text{mag,add}}$ free parameters when not set to zero. Optionally, $\Gamma_{\text{ab,add}}$ and $\Gamma_{\text{em,add}}$ can be reduced into a single parameter through the detailed balance:

\begin{align}
\Gamma_{\text{ab,add}}=\Gamma_{\text{em,add}}\text{exp}\left(-(u_+-u_-)/k_{\text{B}}T\right)
\label{detbalcond}
\end{align}

$\Gamma_{\text{ab,add}}$, $\Gamma_{\text{em,add}}$, $\Gamma_{\text{mag,add}}$ may still have an alternative role: if a different spin Hamiltonian $H_d$ is preferred over Eq.\eqref{spinH}, one would set $\Gamma_{\text{ab}}=\Gamma_{\text{em}}=\Gamma_{\text{mag}}=0$ and input in $\va{\mu}_{+-}$, $\Gamma_{\text{ab,add}}$, $\Gamma_{\text{em,add}}$, $\Gamma_{\text{mag,add}}$ the values calculated with the energy scheme of $H_d$ which now provides $\ket{u_-}, \ket{u_+}, u_-, u_+$. An instance could be the expansion of Eq.\eqref{spinH} with $J_{\text{ex}}$.

\section{Results}

\subsection{Quantum gates for one-qubit algorithms}

\begin{figure}[H]
    \centering
    \includegraphics[scale=0.60]{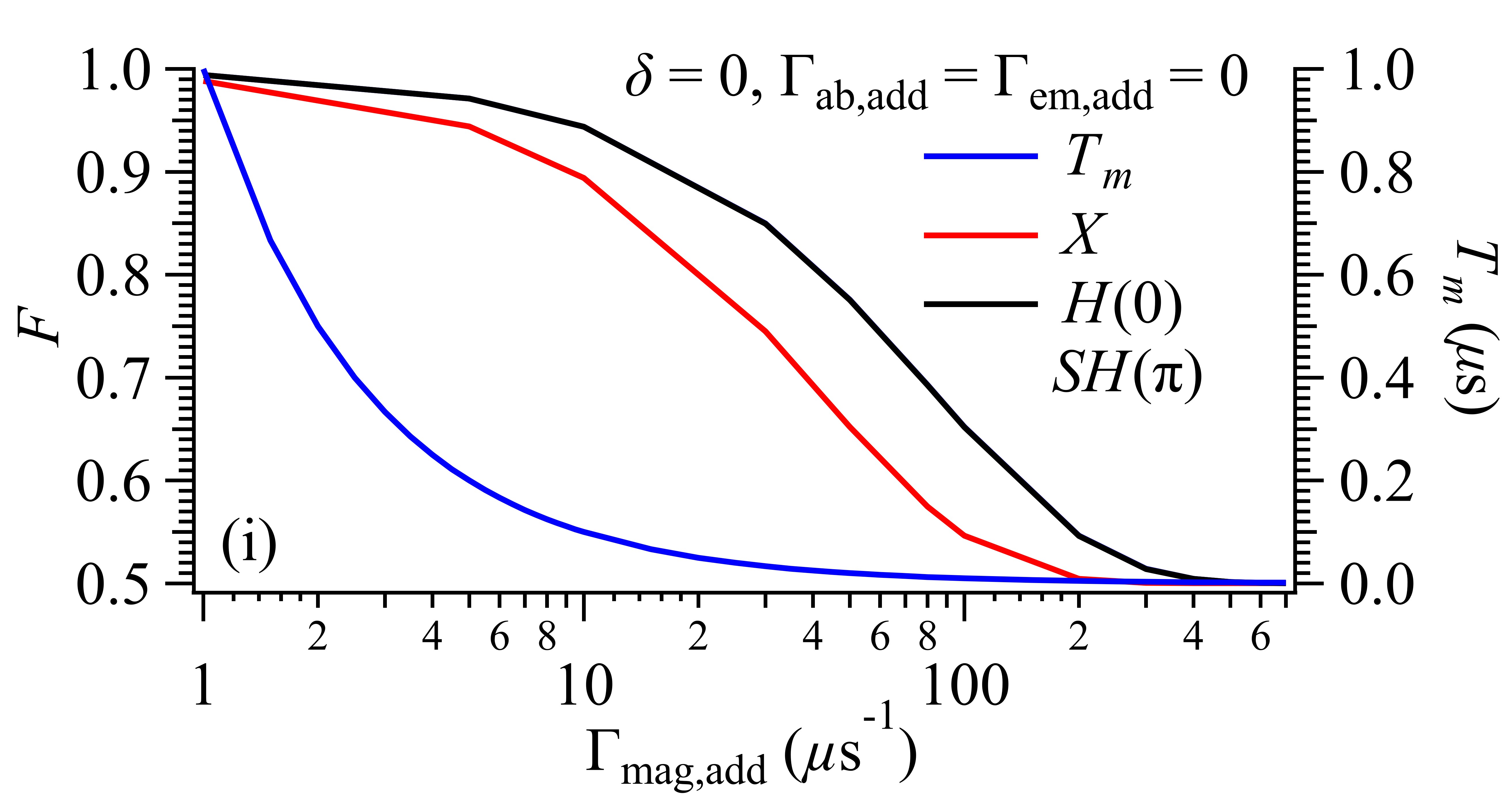}
    \includegraphics[scale=0.60]{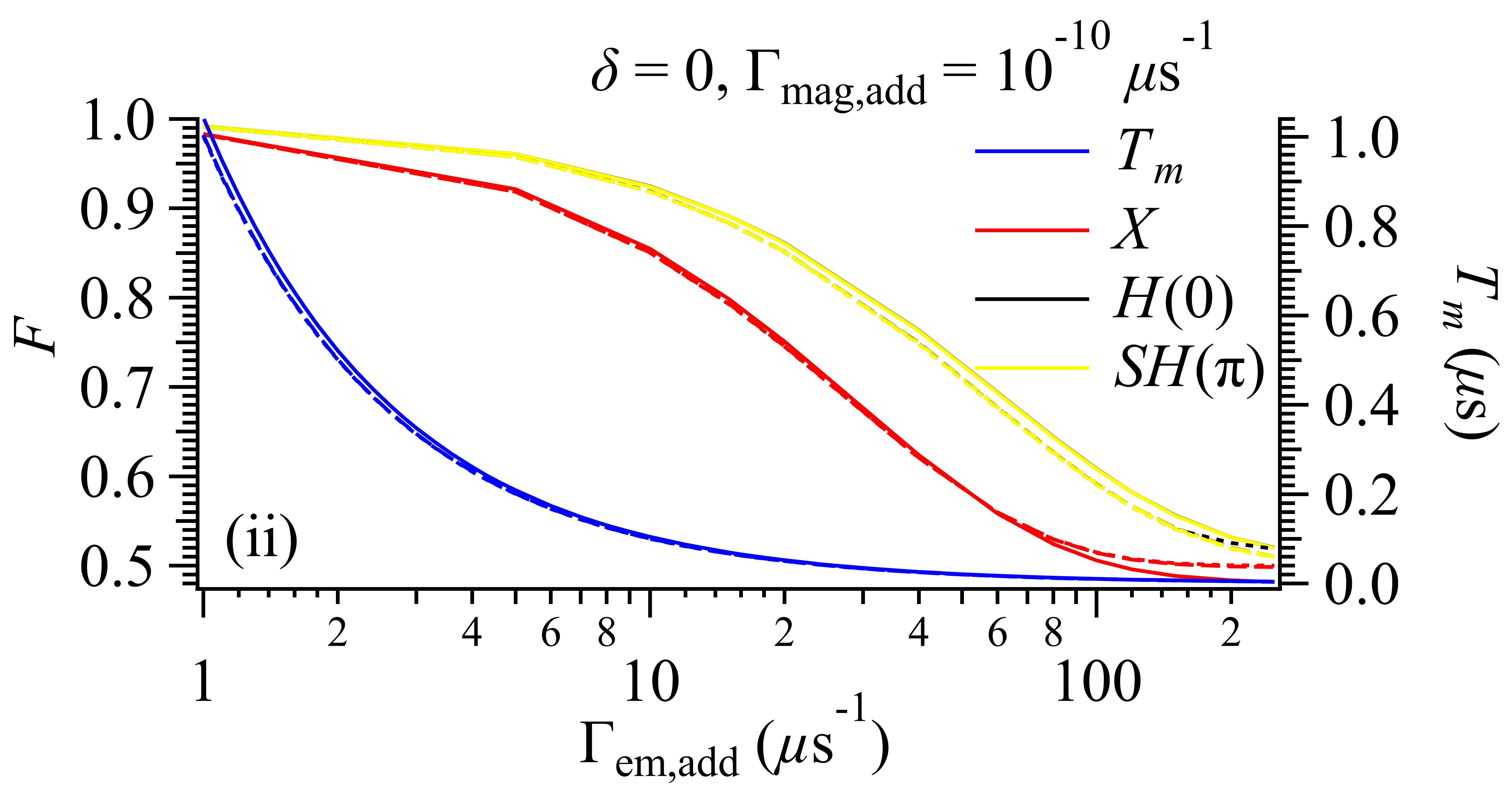}
    \includegraphics[scale=0.60]{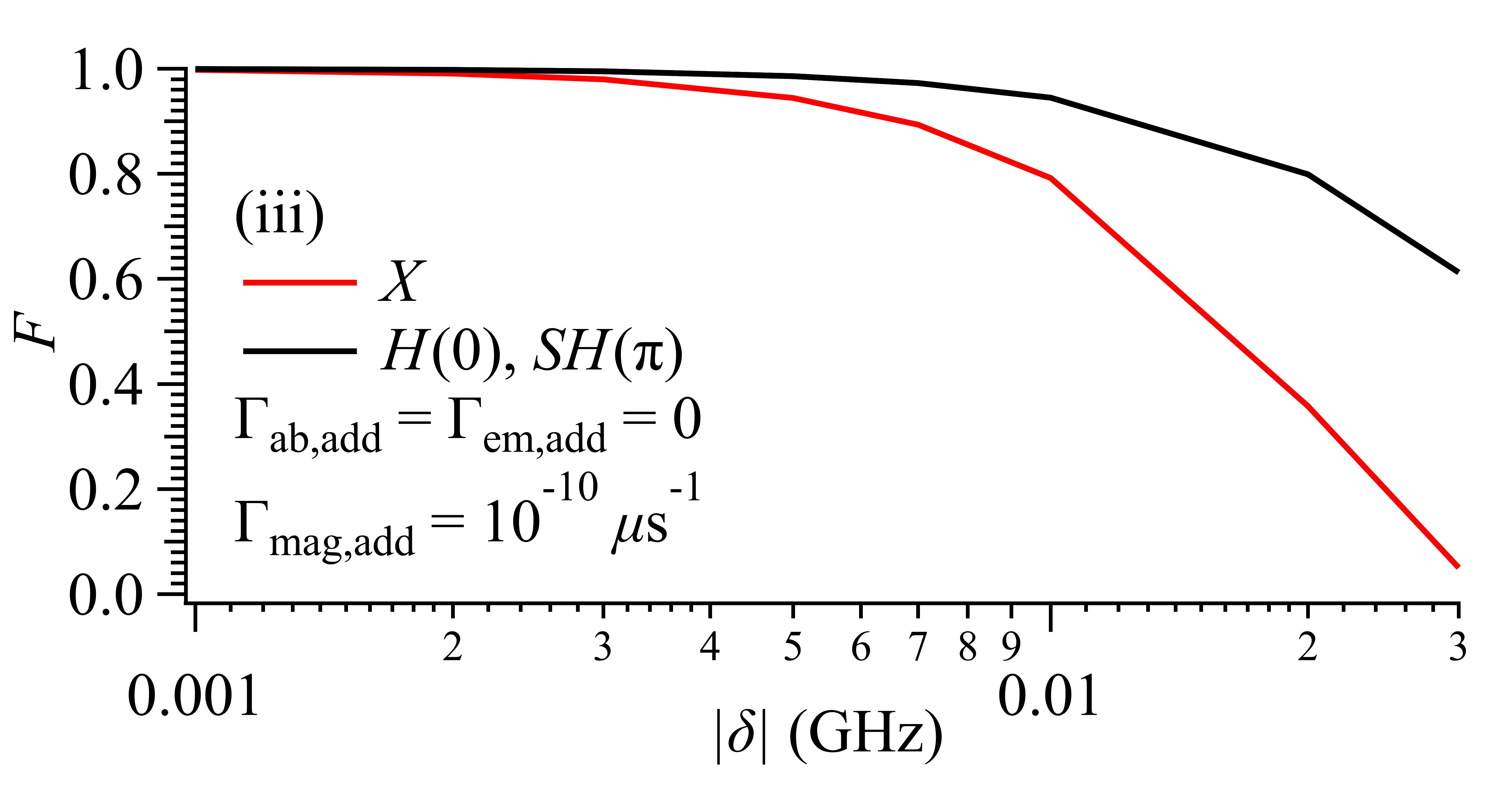}
    \includegraphics[scale=0.60]{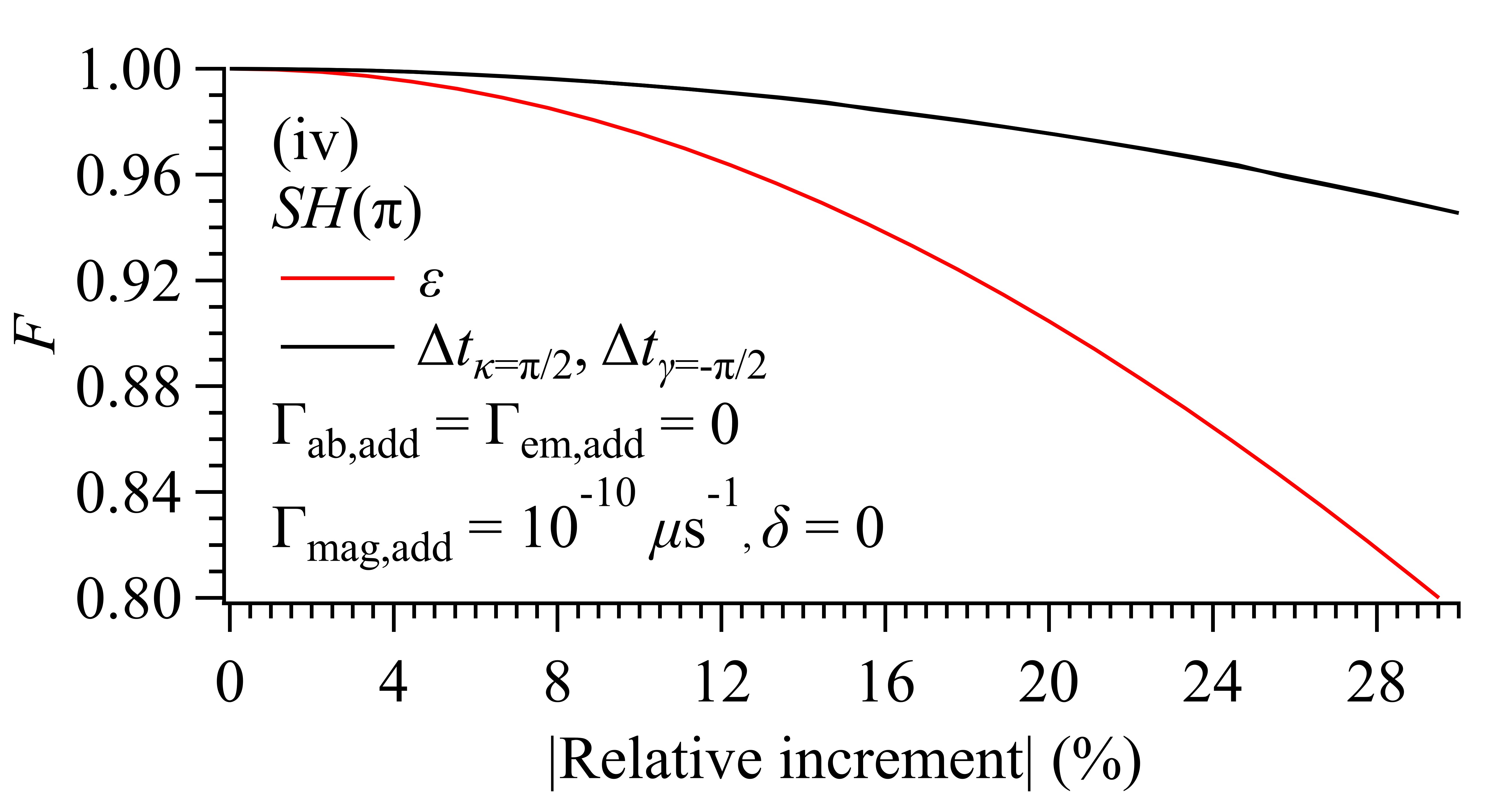}
   \caption{Computed fidelity $F$ of $X$, $H(0)$, $SH(\pi)$ vs (i) $\Gamma_{\text{mag,add}}$, (ii) $\Gamma_{\text{em,add}}$, (iii) $\delta$, and (iv) deviations in $\epsilon$, $\Delta t_{\kappa=\pi/2}$, $\Delta t_{\gamma=-\pi/2}$ relative to $\pi$, $11.9$ ns, $0.028$ ns, resp.; with $\Gamma_{\text{ab}}=\Gamma_{\text{em}}=\Gamma_{\text{mag}}=0$. (i) and (ii) also show the computed $T_m$. Solid, dashed, dotted lines in (ii): 5, 50, 200 K, resp. For numerical reasons, at least one rate ($\Gamma_{\text{mag,add}}$) must be given an effective zero value.}
   \label{fidelidades}
\end{figure}

Rotation and free evolution can be combined in QBithm to control at will the value of $\theta$, $\phi$ and place the spin qubit at any point of the Bloch sphere. Given initial $(\theta_0,\phi_0)$, any other $(\theta_f,\phi_f)$ with $0\leq\theta_0,\theta_f\leq\pi$, $0\leq\phi_0,\phi_f<2\pi$ can be reached as follows. The first step is a rotation of angle $\kappa=\theta_f-\theta_0$. This rotation is performed around the axis -contained in the equatorial plane of the Bloch sphere- whose direction forms an angle $\epsilon=\pi-\phi_0$ respect to the positive $\mathbb{Y}^+$ axis of the said sphere, and takes $\Delta t_{\kappa}=\kappa/\Omega_{\text{g}}$ if $\kappa>0$, or $\Delta t_{\kappa}=(2\pi+\kappa)/\Omega_{\text{g}}$ if $\kappa<0$. Then, given $\gamma=\phi_f-\phi_0$, depending on whether $\gamma<0$ or $\gamma>0$, a free evolution is implemented with $\Delta t_{\gamma}=-\gamma/\omega_{+-}$ or $\Delta t_{\gamma}=(2\pi-\gamma)/\omega_{+-}$. In case of operating with a single rotation axis, e.g. $\epsilon=0$ ($\mathbb{Y}^+$ axis), one first implements a free evolution to reach $(\theta_0,\phi=\pi)$. Then, a rotation around $\epsilon=0$ gets $(\theta_f,\phi=\pi)$. Last, a new free evolution changes $\phi=\pi$ into $\phi_f$.
\medbreak
We have employed QBithm to implement a selection of one-qubit gates key in quantum algorithms: NOT ($N(\epsilon)$), Hadamard ($H(\epsilon)$), Phase ($P(\gamma)$). Note that rotations implemented by our code are clock-wise. The spin qubit is initialized as $\ket{\psi}=\ket{u_-}\equiv\ket{0}$. $N(\epsilon)$, $H(\epsilon)$ are implemented as a rotation gate, and produce clock-wise zenithal rotations in the Bloch sphere with respective angles $\pi$, $\pi/2$ around the axis $\epsilon$ and along the meridian $\phi=\pi-\epsilon$. $P(\gamma)$ is implemented as a free evolution gate, and performs a clock-wise azimuthal rotation around $\mathbb{Z}^+$ with an angle $\gamma<0$ and $\theta$ unaltered. The gates $G$ whose implementation $G\ket{\psi}$ we demonstrate are the Pauli $X=N(\epsilon=0)$, $Y=N(\epsilon=\pi/2)$ gates, $H(\epsilon=\pi/2)$, $H(\epsilon=0)$, $S=P(\gamma=-\pi/2)$, $T=P(\gamma=-\pi/4)$. Since $P(\gamma)\ket{0}=\ket{0}$, to show how $P(\gamma)$ can change $\phi$, it must be applied on states with $\theta\neq 0,\pi$ (namely $\ket{0}$, $\ket{1}$). For $S$, $T$, we choose $\ket{\psi}=H(\epsilon=\pi)\ket{0}=(\ket{1}+\ket{0})/\sqrt{2}$ ($\theta=\pi/2$).
\medbreak
We consider the illustrative case of an effective spin doublet as a spin qubit with $\omega_{+-}/2\pi=0.3$ cm$^{-1}$ $=9$ GHz, $|\va{B}_1|=1.5$ mT, $\delta=0$. This results in $\Omega_{\text{g}}/2\pi=21$ MHz, $\Delta t_{\kappa=\pi}=23.8$ ns, $\Delta t_{\kappa=\pi/2}=11.9$ ns, $\Delta t_{\gamma=-\pi/2}=0.028$ ns, $\Delta t_{\gamma=-\pi/4}=0.014$ ns. With these data as an input, the expected $2\times2$ density matrices $\overline{\rho}=G\ket{\psi}\bra{\psi}G^{\dagger}$ are found to match with those provided by QBithm (see SI).
\medbreak
The gates have been implemented under ideal circumstances: $\Gamma_{\text{ab}}=\Gamma_{\text{em}}=\Gamma_{\text{mag}}=\Gamma_{\text{ab,add}}=\Gamma_{\text{em,add}}=\Gamma_{\text{mag,add}}=0$, $\delta=0$, and the right values of $\epsilon$, $\Delta t_{\kappa}$, $\Delta t_{\gamma}$, thus providing a fidelity $F=1$. We now study how relaxation and imperfections affect $F$. We repeat the same examples, also with $\Gamma_{\text{ab}}=\Gamma_{\text{em}}=\Gamma_{\text{mag}}=0$, but now (i) $\delta=0$, $\Gamma_{\text{ab,add}}=\Gamma_{\text{em,add}}=0$, $\Gamma_{\text{mag,add}}>0$, (ii) $\delta=0$, $\Gamma_{\text{mag,add}}=0$, $\Gamma_{\text{em,add}}>0$ with $T$ = 5, 50, 200 K and $\Gamma_{\text{ab,add}}$ given by Eq.\eqref{detbalcond}, (iii) $\delta>0$, $\Gamma_{\text{ab,add}}=\Gamma_{\text{em,add}}=\Gamma_{\text{mag,add}}=0$, (iv) case of the $SH(\pi)$ gate sequence with $\delta=0$, $\Gamma_{\text{ab,add}}=\Gamma_{\text{em,add}}=\Gamma_{\text{mag,add}}=0$, and with deviations relative to the right values $\epsilon=\pi$, $\Delta t_{\kappa=\pi/2}= 11.9$ ns, $\Delta t_{\gamma=-\pi/2}= 0.028$ ns. For each value of (i) $\Gamma_{\text{mag,add}}>0$, (ii) $\Gamma_{\text{em,add}}>0$, (iii) $\delta>0$, (iv) the relative increment in $\epsilon$, $\Delta t_{\kappa=\pi/2}$, $\Delta t_{\gamma=-\pi/2}$, QBithm delivers a density matrix $\overline{\sigma}$ which we use to evaluate the fidelity $0\leq F(\overline{\rho},\overline{\sigma})\leq 1$ shown in Figure~\ref{fidelidades}. The qubit performance can be equally assessed in any one-spin-qubit algorithm by calculating the fidelity between the ideal and non-ideal density matrices obtained after running the same gate sequence.

In (i), (ii), where relaxation is driven only by the spin bath and only by the vibration bath resp., we also compute $T_m$ to check how it compares with $F$. High fidelities $>$95\% require $T_m \gtrsim 0.89$ $\mu s$ in (i) and $\gtrsim 0.94$ $\mu s$ in (ii) with no imperfections. The little affection by $T$ in (ii) is due to $u_+-u_-=0.3$ cm$^{-1}=0.43$ K in Eq.\eqref{detbalcond} is small enough compared to the explored $T=$ 5, 50, 200 K. While $T_m \sim 1$ $\mu s$ values are achievable by current spin qubits, the interest of our method is here found at determining the needed $T_m$ to drive the qubit through the desired number of gates with the prescribed algorithm fidelity.
\medbreak
We highlight the role of $\delta$, $\epsilon$, $\Delta t_{\kappa}$, $\Delta t_{\gamma}$ as key design parameters since imperfections in the form of deviations from $\delta=0$ and from the right values of $\epsilon$, $\Delta t_{\kappa}$, $\Delta t_{\gamma}$ further lower $F$. In (iii), a longer $\va{B}_1$-driven rotation -like that producing $X$ respect to the one involved in $H(0)$, $SH(\pi)$- leads to a lower $F$ for a fixed $\delta$ and requires a smaller $|\delta|$ to achieve a given $F$. $F>95$\% is found for $|\delta|<5$ MHz in $X$ and $|\delta|<10$ MHz in $H(0)$, $SH(\pi)$ if relaxation is negligible. On its hand, (iv) shows that fine-tuning of $\epsilon$ requires more attention than that of $\Delta t_{\kappa}$, $\Delta t_{\gamma}$. The relative increment in $\epsilon$ must not exceed 14.5\% to keep $F>95$\%, while this $F$ can already be obtained with relative increments in $\Delta t_{\kappa}$, $\Delta t_{\gamma}$ below 28.5\%. Note that $H(0)$, $SH(\pi)$ only differ in a free evolution. Since its runtime $0.028$ ns is shorter enough than the explored $\Gamma_{\text{mag,add}}^{-1}$, $\Gamma_{\text{em,add}}^{-1}$ values and $\delta$ is always equal to the constant value $\omega_{+-}$ in any free evolution ($\omega_{\text{MW}}=0$), the $F$ of $H(0)$, $SH(\pi)$ is virtually the same in (i), (ii), (iii).

\subsection{Pulse sequences: Rabi oscillations and spin relaxation times}

We now use QBithm to implement well-known pulse sequences, and reproduce the experimental Rabi oscillations and spin relaxation times $T_1$, $T_m$, $T_{dd}$ of \textbf{(1)}-\textbf{(4)}. Our focus is Eq.\eqref{mastereq}, hence we leave aside the separated task, addressed over the last decade, of calculating $\Gamma_{\text{ab}}$, $\Gamma_{\text{em}}$.\cite{espvib20173,espvib2018,espvib2019,espvib20211,espvib20212,espvib20223,espvib20224,espvib2023,espvib20171,espvib20172,espvib2020,espvib20221,Maestra20194} We set $\Gamma_{\text{ab}}$=$\Gamma_{\text{em}}$=$\Gamma_{\text{mag,add}}$=0, employ the computed $\Gamma_{\text{mag}}$ according to Ref.~\citenum{espesp2016,espesp2019}, and operate QBithm with $\ket{\psi}=\ket{u_-}$ as the initial qubit state and $\Gamma_{\text{em,add}}$ as the only free parameter. Eq.\eqref{detbalcond} provides $\Gamma_{\text{ab,add}}$. Eq.\eqref{spinH} is diagonalized by SIMPRE,\cite{espesp2016} with $\omega_{+-}$ resulting in $\sim 9.6-9.7$ GHz.
\medbreak

\begin{figure*}[t]
    \includegraphics[width=\textwidth]{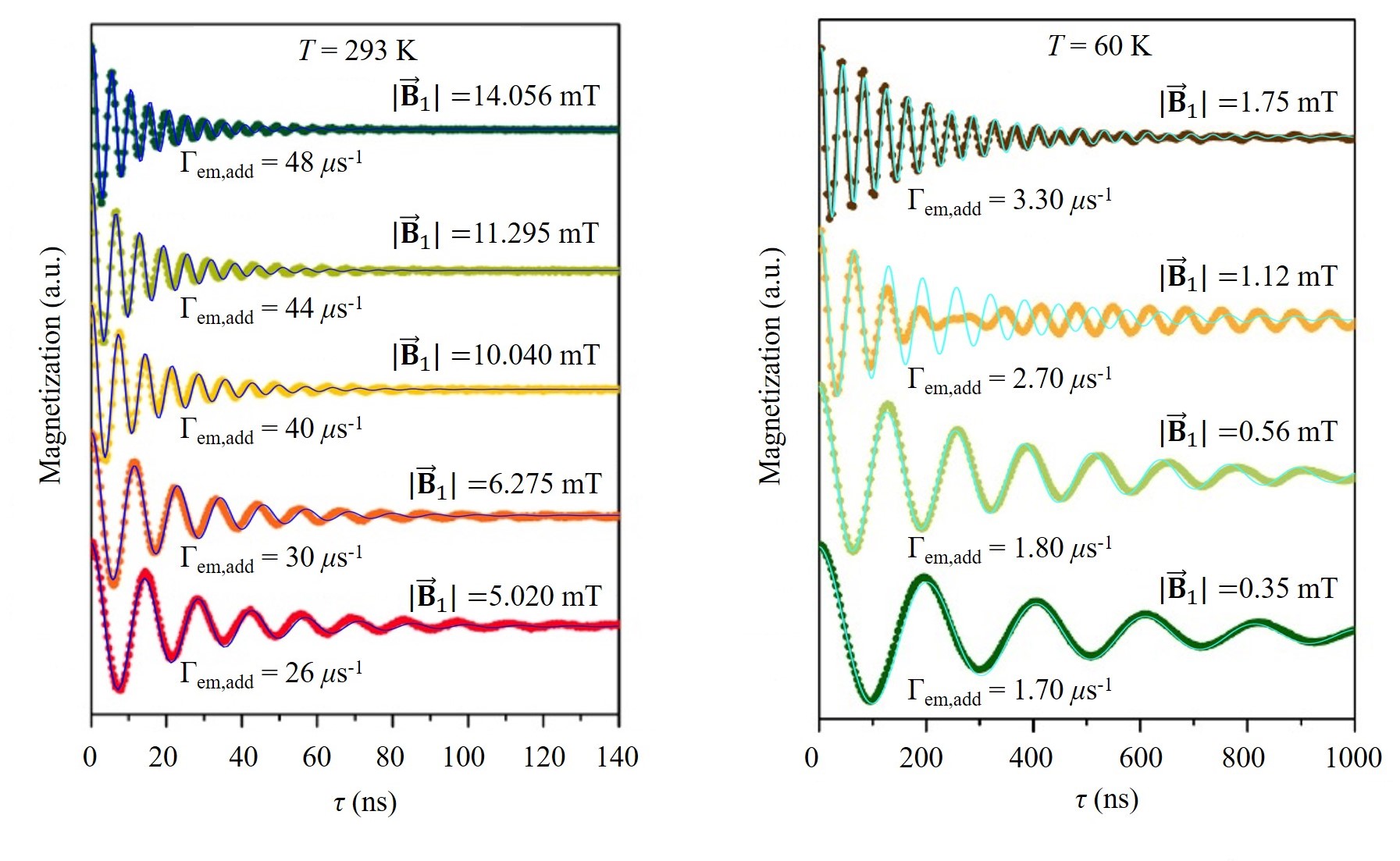}
    \caption{Experimental (dotted \cite{Sistemas20162}) and calculated $\langle\text{M}_z\rangle(t_1)$ (solid) magnetization vs nutation time $\tau$, together with the $\Gamma_{\text{em,add}}$ and $|\va{B}_1|$ values used in QBithm. Left: \textbf{(1)}. Right: \textbf{(2)}.}
    \label{RabiVOdmit2Vdmit3}
\end{figure*}

\paragraph{Rabi oscillations} The sequence to produce Rabi oscillations contains a single gate $n=1$: a rotation -or nutation- around an axis contained in the equatorial plane of the Bloch sphere with a variable $\tau=\Delta t_1$. The plot to perform is $\langle\text{M}_z\rangle(t_1)$ vs $\tau$. In each case study, we find $|\va{B}_1|_1$ of the experimental oscillation determined at the highest attenuation by matching the calculated and the experimental oscillation period $P_1$. Being $|\va{B}_1|_{\text{i,rel}}$ the experimental relative magnitudes,\cite{Sistemas20162,Sistemas20161,Sistemas2014} $|\va{B}_1|_{\text{i}}=|\va{B}_1|_{\text{i,rel}}|\va{B}_1|_1$ for the remaining oscillations. Then, we adjust $\Gamma_{\text{em,add}}$ to reproduce the time decay of each experimental oscillation. See Figure~\ref{RabiVOdmit2Vdmit3} for \textbf{(1)}, \textbf{(2)}, and SI for \textbf{(3)}, \textbf{(4)}.
\medbreak
In Figure~\ref{RabiVOdmit2Vdmit3} Right, \textbf{(2)} displays the onset of a long-lived experimental oscillation at $\tau\sim 200$ ns with a well-defined $\Omega_{\text{HH,g}}^{\text{exp}}$ when $|\va{B}_1|=1.12$ mT. According to the Fourier transform,\cite{Sistemas20162} $\Omega_{\text{HH,g}}^{\text{exp}}/2\pi=14.68$ MHz which coincides with the proton $^1\text{H}$ Larmor frequency $\omega_{^1\text{H}}$ at the working $|\va{B}|=345$ mT. The requirement for the onset of this long-lived oscillation -Hartmann-Hahn condition- is two-fold: a nearby proton hyperfine-coupled to \textbf{(2)}, and to drive \textbf{(2)} with the attenuation $A_{\text{MW}}$ that makes $\Omega_{\text{g}}^{\text{exp}}$ match $\omega_{^1\text{H}}$ at the given $|\va{B}|$. Our theoretical model does not consider any explicit proton, hence it is unable to reproduce the said oscillation.
\medbreak
Note that the sole decrease of $A_{\text{MW}}$ also produces a faster decay of the experimental Rabi oscillations. Here, our values of $\Gamma_{\text{em,add}}$ must be split into the constant contribution $\Gamma_{\text{vb}}$ from the vibration bath and that from an additional rate $\Gamma_{\text{MW}}$ that grows up as $A_{\text{MW}}$ is reduced to recover the mentioned faster decay. The appropriate modeling of this fact, attributed to the static fluctuation of the microwave power,\cite{Sistemas2017} would require a modification of Eq.\eqref{mastereq} to recover $\Gamma_{\text{MW}}$ with $|\va{B}_1|\neq 0$ and $\Gamma_{\text{em,add}}=\Gamma_{\text{vb}}$.

\paragraph{$T_1$ and $T_m$} The Inversion Recovery sequence (IRS) to determine $T_1$ consists of $n=2$ gates being $G_1$ a $\pi$ rotation transforming the initial $\ket{u_-}$ into $\ket{u_+}$, and $G_2$ a free evolution with a variable $\tau=\Delta t_2$. $T_m$ is determined via the Hahn sequence (HS) with $n=4$ gates: $G_1$ a $\pi/2$ rotation producing an equally-weighted superposition between $\ket{u_-}$ and $\ket{u_+}$ from $\ket{u_-}$, $G_3$ a $\pi$ rotation, and $G_2$, $G_4$ free evolutions with the same variable $\tau=\Delta t_2=\Delta t_4$. The fit of the plots $\langle\text{M}_z\rangle(t_2)$ vs $\tau$, $|\langle\text{M}_{xy}\rangle(t_4)|$ vs $2\tau$ to a decaying exponential function provides $T_1$, $T_m$, resp. We model each sequence, with given working conditions, as a unique physical process characterized by a specific set of values $\{\Gamma_{\text{ab}},\Gamma_{\text{em}},\Gamma_{\text{mag}},\Gamma_{\text{ab,add}},\Gamma_{\text{em,add}},\Gamma_{\text{mag,add}}\}$. Our operation mode sets $\Gamma_{\text{ab,em}}^{\text{Rabi}/T_1/T_m/T_{dd}}=0$, $\Gamma_{\text{mag,add}}^{\text{Rabi}/T_m}=0$, and binds $\Gamma_{\text{ab,add}}^{\text{Rabi}/T_1/T_m/T_{dd}}$ with $\Gamma_{\text{em,add}}^{\text{Rabi}/T_1/T_m/T_{dd}}$ via Eq.\eqref{detbalcond}. 
\medbreak
The experimental $T_1$ are reproduced by invoking only the vibration bath with $\Gamma_{\text{mag}}^{T_1}=\Gamma_{\text{mag,add}}^{T_1}=0$ and $\Gamma_{\text{em,add}}^{T_1}$ as the only free parameter. If $T_1$ of a case study different from ours was total or partially limited by the spin bath, the relevant contribution would be added to $\Gamma_{\text{m}}^{T_1}$. We also use $\Gamma_{\text{em,add}}^{T_m}$ as the only free parameter to reproduce the experimental $T_m$. The $\pi/2$ and $\pi$ rotations run by us take $\Delta t_{\pi/2}$, $\Delta t_{\pi}=2\Delta t_{\pi/2}$ with $\Delta t_{\pi/2}=16$ ns for \textbf{(1)}, \textbf{(2)} as experimentally implemented.\cite{Sistemas20162} Now, since $|\va{B}_1|_1P_1$ is a constant and $4\Delta t_{\pi/2}$ is the rotation period, we use $P_1|\va{B}_1|_1=4\Delta t_{\pi/2}|\va{B}_1|_{\text{S}}$ to obtain the value $|\va{B}_1|_{\text{S}}$ to be employed in IRS and HS. This method provides $|\va{B}_1|_{\text{S}}=1.18,$ $1.09$ mT for \textbf{(1)}, \textbf{(2)} resp. The similar case of \textbf{(3)}, \textbf{(4)} is discussed in SI.
\medbreak
The fit of the experimental $\langle\text{M}_z\rangle$ of \textbf{(1)}, \textbf{(2)} to $\propto\text{exp}(-(\tau/T_1)^x)$ delivers rather low stretch factors $0.4<x<0.8$.\cite{Sistemas20162} According to Ref.~\citenum{Sistemas2014,millicohtime}, $\langle\text{M}_z\rangle$ can be interpreted as the result of a fast and a slow timescale $T_{1,f}$, $T_{1,s}$ with the former operating at short times and the latter at longer ones. We thus fit the calculated $\langle\text{M}_z\rangle(t_2)$ vs $\tau$ to $\propto a_f\text{exp}(-\tau/T_{1,f})+a_s\text{exp}(-\tau/T_{1,s})$ and, in $4.5-294$ K, obtain $1.5\cdot10^{-4}\leq\Gamma_{\text{em,add}}^f\leq4.3\cdot10^{-1}$ $\mu$s$^{-1}$, $2.0\cdot10^{-5}\leq\Gamma_{\text{em,add}}^s\leq1.6\cdot10^{-1}$ $\mu$s$^{-1}$ for \textbf{(1)}, and $3.4\cdot10^{-3}\leq\Gamma_{\text{em,add}}^f\leq7.8\cdot10^{-1}$ $\mu$s$^{-1}$, $2.8\cdot10^{-4}\leq\Gamma_{\text{em,add}}^s\leq7.4\cdot10^{-1}$ $\mu$s$^{-1}$ for \textbf{(2)}. The fit of the calculated $|\langle\text{M}_{xy}\rangle(t_4)|$ vs $2\tau$ to $\propto\text{exp}(-(2\tau/T_m)^x)$ produces $0.95\geq x\geq 0.92$, $0.75\leq\Gamma_{\text{em,add}}\leq1.82$ $\mu$s$^{-1}$ for \textbf{(1)} and $x=1.0$, $1.91\leq\Gamma_{\text{em,add}}\leq7.01$ $\mu$s$^{-1}$ for \textbf{(2)} in $4.5-294$ K.

\paragraph{CPMG sequence} This sequence is based on the block $\{G_2\rightarrow G_3\rightarrow G_2\}$ which is repeated $k$ times after an initial $\pi/2$ rotation $G_1$ around $\mathbb{X}^+$. $G_2$ is a free evolution with a variable $\tau=\Delta t_2$, while $G_3$ is a $\pi$ rotation around $\mathbb{Y}^+$. The number of gates is $n=3k+1$ and $2k$ of them are run for $\tau$. Given $k$, $T_{dd}$ is extracted from fitting $|\langle\text{M}_{xy}\rangle(t_{3k+1})|$ vs $2k\tau$ to $\propto\text{exp}\left(-(2k\tau/T_{dd})^x\right)$. The key to CPMG sequence is the continual implementation of the said block which mitigates the relaxation produced by the spin bath and lengthen $T_{dd}$ as $k$ is increased.\cite{Sistemas2017} In practice, $T_{dd}$ is limited by experimental imperfections.

\begin{figure}[H]
    \centering
    \includegraphics[scale=0.60]{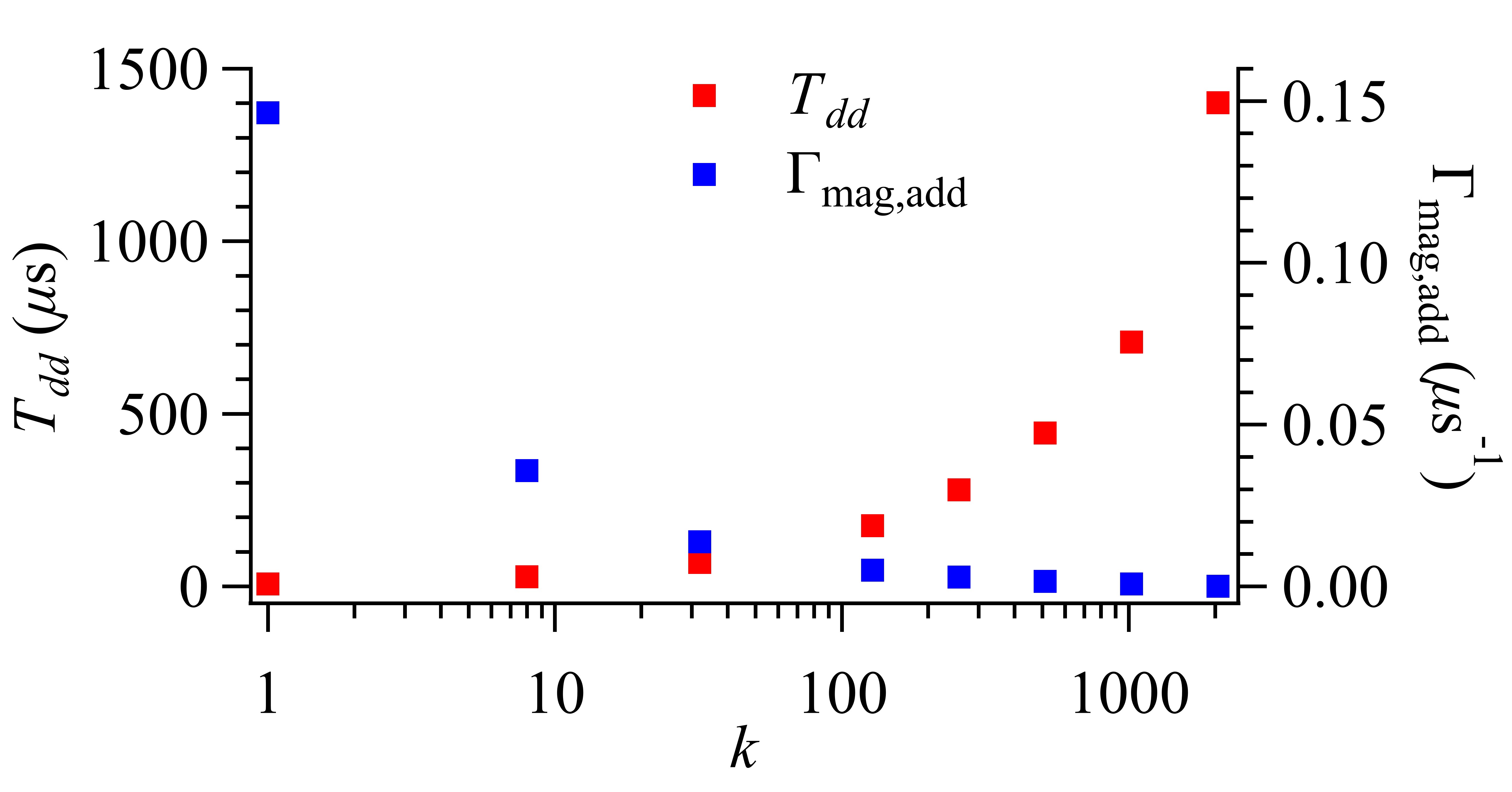}
    \includegraphics[scale=0.60]{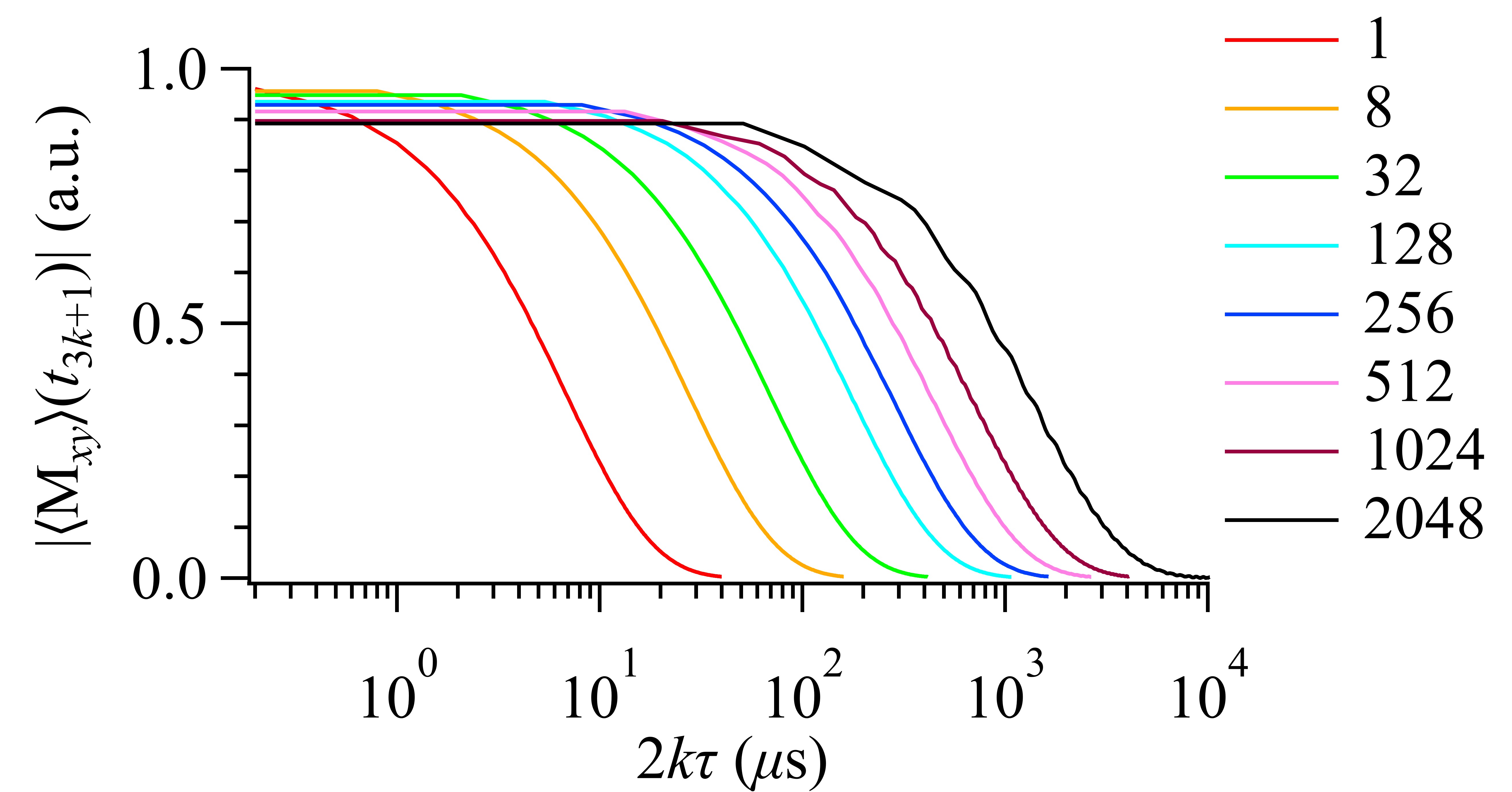}
   \caption{Top: $\Gamma_{\text{mag,add}}$ rates determined with QBithm to fit the experimental relaxation times $T_{dd}$ measured when applying the CPMG sequence to \textbf{(4)} with $k=1-2048$.\cite{Sistemas2017} Bottom: computed in-plane magnetization curves at each $k$.}
   \label{CPMGCumnt2Jiangfeng}
\end{figure}

The CPMG sequence is run at the fixed $T=8$ K and, since its role is to essentially quench the dephasing produced by the spin bath, we model the $T_{dd}$ lengthening as a reduction in $\Gamma_{\text{mag,add}}^{T_{dd}}$ with a constant $\Gamma_{\text{em,add}}^{T_{dd}}=\Gamma_{\text{em,add}}^{T_{dd},\text{lim}}$. $\Gamma_{\text{mag}}^{T_{dd}}$ is set to zero thus the entire relaxation due to the spin bath is collected by $\Gamma_{\text{mag,add}}^{T_{dd}}$ as the only free parameter. We determine $\Gamma_{\text{em,add}}^{T_{dd},\text{lim}}$ by fitting the calculated $T_{dd}$ to the limiting value $T_{dd}^{\text{lim}}$ measured at a high enough $k_{\text{lim}}$. Since the qubit is already certainly decoupled from the spin bath at $k=k_{\text{lim}}$, we employ $\Gamma_{\text{mag,add}}^{T_{dd}}=0$ in this particular fitting process. Then, for each $k<k_{\text{lim}}$ explored, we keep $\Gamma_{\text{em,add}}^{T_{dd}}=\Gamma_{\text{em,add}}^{T_{dd},\text{lim}}$ and vary $\Gamma_{\text{mag,add}}^{T_{dd}}$ until matching the experimental $T_{dd}$. We study the case of \textbf{(4)}, where we lack $T_{dd}^{\text{lim}}$. Hence, to illustrate our method, we use $T_{dd}^{\text{lim}}=1.4$ ms obtained with the largest $k=2048$ implemented. We use the experimental $\Delta t_{\pi/2}=24$ ns, $\Delta t_{\pi}=48$ ns which provide $|\va{B}_1|_{\text{DD}}=0.76$ mT, and get $\Gamma_{\text{em,add}}^{T_{dd},\text{lim}}=7.3\cdot 10^{-4}$ $\mu$s$^{-1}$, $x=1.0$, $1.46\cdot10^{-1}\geq\Gamma_{\text{mag,add}}^{T_{dd}}\geq 0$ $\mu$s$^{-1}$ for $k=1-2048$, see Figure~\ref{CPMGCumnt2Jiangfeng} and SI.

\section{Discussion}

Over the last decade, the study of $T_1$ and $T_m$ has undoubtedly produced a body of design principles aimed at quenching the loss of quantum information stored in freely-evolving spin qubits. At this point, these well-established strategies that have resulted from a strong interplay theory-experiment set an ideal position for pushing the use of spin qubits in quantum algorithms as a key application. This means to submit the qubit to an evolution not free anymore but rather driven by the user while relaxation and imperfections are active. In this widened picture, $T_1$ and $T_m$ provide a limited benchmark for the qubit performance which must now be evaluated after running the given algorithm in the form of fidelity as the actual figure of merit thus accounting for not only relaxation but also for other sources of error.
\medbreak
We have posed and analytically solved a master equation purposely designed to run any one-spin-qubit algorithm driven by the user as a gate sequence subject to relaxation and potential experimental imperfections. The fidelity $F$ is easily computed by combining the two qubit density matrices obtained after running the algorithm with and without relaxation rates $\Gamma$ and imperfections. This allows assessing a qubit among several algorithms as well as comparing different qubits each one implementing the same algorithm. While other approaches offer a rotation angle $\zeta$ as the only input,\cite{Maestra20191} our framework delivers a higher degree of control by letting the user input $|\va{B}_1|$, $\omega_{\text{MW}}$, gate time $\Delta t_i$ -which all make up $\zeta$- and the rotation axis $\epsilon$. This allows (i) a more realistic description of rotations with $\delta\neq0$, $\Gamma\neq0$, and (ii) to study the impact of deviations in ideal values of the said parameters. 
\medbreak
The master equation is operated through the open-source home-made code QBithm which, in addition to show the time evolution of the density matrix, also offers the possibility of calculating longitudinal and in-plane magnetization curves to determine Rabi oscillations, relaxation times such as $T_1$, $T_m$, and ESE-detected spectra. This feature helps to translate between $T_m$ and $F$ in a given algorithm: one can learn both what $F$ would result from the $\Gamma$ values reproducing $T_m$, and which the minimum $T_m$ should be so the $\Gamma$ values lead to a prescribed $F$.
\medbreak
The operation mode in QBithm depends on the $\Gamma$ values whether they are previously calculated or employed as free parameters. In the first case, our work takes up the torch from the \textit{ab initio} methods devoted to compute the rates $\Gamma$ which now feed QBithm. If using $\Gamma$ to fit experimental data, the resulting values can serve as a benchmark for the mentioned methods. Here, we have operated under a combined mode that allows dealing with a plethora of potential case studies and experiments of wide interest. We managed to reproduce the experimental data by describing the complex issue of relaxation with realistic values of a single relaxation rate as the only free parameter. At this point, the ball is now in the court of those researchers interested in proceeding through a purely \textit{ab initio} mode.
\medbreak
We did compute $\Gamma_{\text{ab}}$ and $\Gamma_{\text{em}}$ for \textbf{(4)} from its calculated molecular-vibration spectrum.\cite{espvib20171}. The obtained values $<10^{-10}$ $\mu$s$^{-1}$ in $5-300$ K are rather negligible and cannot account for the experimental $T_1$ and $T_m$ lying in $1$ $\mu s$ $-$ $100$ $\text{ms}$. We find that key to such a low values is the mismatch between the spin-qubit gap $\omega_{+-}\sim0.3$, $\sim1.0$ cm$^{-1}$ -corresponding to the X and Q EPR-pulse bands- and the harmonic frequencies $\omega_i\gtrsim 20$ cm$^{-1}$ of the molecular vibrations, as well as the low phonon populations that $\omega_i$ produce at the working $T$. Hence, we state that calculated vibration spectra should also include the low-energy lattice vibrations -specially those with $\omega_i\lesssim 1$ cm$^{-1}$- for a proper description of the direct relaxation process between the two qubit states $\ket{u_+}$ and $\ket{u_-}$.\cite{Maestra20194,Maestra20222} Indeed, after resetting $\omega_i$ to 1 cm$^{-1}$ and by assuming that both spin-vibration matrix elements and reduced masses remain unaltered, we found that each molecular vibration contributes to $\Gamma_{\text{ab}}$, $\Gamma_{\text{em}}$ with values $\sim 10^6-10^8$ times larger in $5-300$ K. Hence, if $\sim 10-100$ lattice vibrations with $\omega_i\lesssim 1$ cm$^{-1}$ were included, $\Gamma_{\text{ab}}$ and $\Gamma_{\text{em}}$ would increase by $\sim 10^7-10^{10}$ thus resulting in more realistic values in the $\gtrsim\mu$s scale.
\medbreak
Molecular/local vibrations should not, however, be discarded as they can play a significant role in Raman relaxation.\cite{Maestra20221} Given two vibration modes $i$, $j$, the mismatch of $\omega_i$, $\omega_j$ respect to $\omega_{+-}$ is not crucial since the resonance condition to fulfil now is $|\omega_i-\omega_j|\sim\omega_{+-}$. While this condition suffices to account for virtual Raman processes, real Raman processes additionally require an eigenstate $\ket{u_c}$ of Eq.\eqref{spinH} as an intermediate state -with energy either $u_c>u_+$ or $u_c<u_-$- such that $|u_c-u_+|$, $|u_c-u_-|$ must be similar to one among $\omega_i$, $\omega_j$. The sole inclusion in Eq.\eqref{spinH} of the ground $J$, $I$ quantum numbers of \textbf{(1)}-\textbf{(4)} provides energy schemes of $(2J+1)(2I+1)$ eigenstates that span no more than 1 cm$^{-1}$. Hence, $|u_c-u_+|$, $|u_c-u_-|$ $\lesssim 1$ cm$^{-1}$. While this fact also supports the need of incorporating lattice vibrations with $\omega_i\lesssim 1$ cm$^{-1}$, molecular vibrations -which usually have $\omega_i\gtrsim 20$ cm$^{-1}$- will play a significant role only if there exist $\ket{u_c}$ such that $|u_c-u_+|$, $|u_c-u_-|$ $\gtrsim 20$ cm$^{-1}$. Thus, for a proper description of molecular/local-vibration relaxation, we also state that one could require to expand Eq.\eqref{spinH} with $J_{ex}$ or even with other degrees of freedom which also couple to $J$ and produce states that might also work as high-energy intermediate states. As included in our framework, both real and virtual Raman processes can also proceed through intermediate states with energy between $u_-$ and $u_+$, being the resonance condition $\omega_i+\omega_j\sim\omega_{+-}$. Since $\omega_{+-}\lesssim 1$ cm$^{-1}$, the proper modeling of this process would require again the participation of low-energy vibrations.
\medbreak
We stress the role of $\delta$, $\epsilon$, $\Delta t_{\kappa}$, $\Delta t_{\gamma}$ as crucial design parameters to be addressed. Indeed, imperfections leading to deviations from their ideal values also result in lower fidelities apart from a short $T_m$. Ideally, one would operate with $\delta=\omega_{+-}-\omega_{\text{MW}}=0$ but, if $\delta\neq 0$, one can still conduct a faithful driving provided $\omega_{+-}$ lies inside the excitation bandwidth $\text{B}=[\omega_{\text{MW}}-\Delta\omega_{\text{MW}}/2,\omega_{\text{MW}}+\Delta\omega_{\text{MW}}/2]$. For pulses at least 10 ns long like those employed by us, $\Delta\omega_{\text{MW}}\leq 100$ MHz in standard EPR setups. Moreover, the distribution $\text{D}$ of the parameters in Eq.\eqref{spinH} as a result of static disorder leads to a broadening in $\omega_{+-}$ whose FWHM can be as high as 300 MHz.\cite{gapwidth2016} Hence, there exists a window as wide as $\Delta\omega_{\text{MW}}/2+\text{FWHM}/2=200$ MHz where the tails of $\text{B}$ and $\text{D}$ can still overlap provided $|\delta|<200$ MHz. This is the case of \textbf{(1)}-\textbf{(4)}, where the minimum detuning $|\delta|_{\text{min}}$ ranges from 10 to 150 MHz. Instead of including $\Delta\omega_{\text{MW}}$ and FWHM in our model, and since the resonance probability decays abruptly as soon as $|\delta|$ becomes larger, we have opted for a simpler alternative which effectively covers the relevant physics: we use the $|\va{B}|$ direction $\va{u}$ where $|\delta|=|\delta|_{\text{min}}$ and give $\omega_{\text{MW}}$ the value of $\omega_{+-}$ corresponding to $\va{u}$. If there existed several $\va{u}$ directions (isotropic systems), one would integrate the magnetization curve over all $\va{u}$ like in powder samples and frozen solutions (see SI).
\medbreak
QBithm also offers an appealing opportunity for experimentalists as it may help to characterize EPR-pulse experiments in terms of key design parameters mostly gone unheeded so far. These include $\delta$, $|\va{B}_1|$, $\va{B}_1$ polarization, $\epsilon$, and gate times, which can all be tested against deviations from ideal values to check the impact on the qubit performance. Moreover, Eq.\eqref{Heff} can be modified to implement the potential option of driving the spin via an oscillating electric field as already explored.\cite{ElecCont20211} A natural next step is the expansion of the computational space $\{\ket{u_+},\ket{u_-}\}$ to firstly accommodate two spin qubits and then a larger set to run sophisticated algorithms. The former results in the implementation of two-qubit gates to test their fidelity,\cite{Fidel2022} and in the inclusion of physical processes not collected by our model like the interaction between a single qubit and a nearby nuclear spin-1/2, which may be behind long-lived Rabi oscillations and oscillatory behaviors of magnetization curves. 
\medbreak
In a nutshell, our contribution comes with an assorted toolkit that tackles the issue of driving one spin qubit on the Bloch sphere while exposed to relaxation and imperfections, thus supplying the complementary piece of those design principles devised to lengthen $T_m$ for a spin in free evolution. We expect that our work will be of utility to the wide community of spin qubits and to those researchers interested in exploiting this physical platform for quantum information and quantum computation.

\section{Acknowledgements}
The author thanks financial support from European Union (FET-OPEN FATMOLS 862893), Spanish MICIN (Unit of Excellence ‘María de Maeztu’ CEX2019-000919-M), and Generalitat Valenciana (Prometeo program PROMETEO/2021/022). Computation resources have been provided by the scientific calculation cluster ‘Lluis Vives’ of the University of Valencia.

\bibliographystyle{quantum}
\bibliography{quantum-template}

\begin{thebibliography}{10}

\bibitem{Intro20202}
B.~Bauer, S.~Bravyi, M.~Motta, and G.~Kin-Lic Chan.
\newblock ``Quantum algorithms for quantum chemistry and quantum materials science''.
\newblock \href{https://dx.doi.org/https://doi.org/10.1021/acs.chemrev.9b00829}{Chem. Rev. {\bf 120}, 12685–12717}~(2020).

\bibitem{Intro20214}
K.~Head-Marsden, J.~Flick, C.~J. Ciccarino, and P.~Narang.
\newblock ``Quantum information and algorithms for correlated quantum matter''.
\newblock \href{https://dx.doi.org/https://doi.org/10.1021/acs.chemrev.0c00620}{Chem. Rev. {\bf 121}, 3061–3120}~(2021).

\bibitem{Intro20201}
L.~Henriet, L.~Beguin, A.~Signoles, T.~Lahaye, A.~Browaeys, G.-Olivier Reymond, and C.~Jurczak.
\newblock ``Quantum computing with neutral atoms''.
\newblock \href{https://dx.doi.org/https://doi.org/10.22331/q-2020-09-21-327}{Quantum {\bf 4}, 327}~(2020).

\bibitem{Intro20171}
B.~Lekitsch, S.~Weidt, A.~G. Fowler, K.~Mølmer, S.~J. Devitt, C.~Wunderlich, and W.~K. Hensinger.
\newblock ``Blueprint for a microwave trapped ion quantum computer''.
\newblock \href{https://dx.doi.org/https://doi.org/10.1126/sciadv.1601540}{Sci. Adv. {\bf 3}, 2}~(2017).

\bibitem{Intro2019}
X.~Liu and M.~C. Hersam.
\newblock ``2\uppercase{D} materials for quantum information science''.
\newblock \href{https://dx.doi.org/https://doi.org/10.1038/s41578-019-0136-x}{Nat. Rev. Mater. {\bf 4}, 669–684}~(2019).

\bibitem{Intro20211}
A.~J. Heinrich, W.~D. Oliver, L.~M.~K. Vandersypen, A.~Ardavan, R.~Sessoli, D.~Loss, A.~B. Jayich, J.~Fernández-Rossier, A.~Laucht, and A.~Morello.
\newblock ``Quantum-coherent nanoscience''.
\newblock \href{https://dx.doi.org/https://doi.org/10.1038/s41565-021-00994-1}{Nat. Nanotechnol. {\bf 16}, 1318–1329}~(2021).

\bibitem{Review2021}
C.-Jui Yu, S.~von Kugelgen, D.~W. Laorenza, and D.~E. Freedman.
\newblock ``A molecular approach to quantum sensing''.
\newblock \href{https://dx.doi.org/https://doi.org/10.1021/acscentsci.0c00737}{ACS Cent. Sci. {\bf 7}, 712–723}~(2021).

\bibitem{Review2022}
D.~W. Laorenza and D.~E. Freedman.
\newblock ``Could the quantum internet be comprised of molecular spins with tunable optical interfaces?''.
\newblock \href{https://dx.doi.org/https://doi.org/10.1021/jacs.2c07775}{J. Am. Chem. Soc. {\bf 144}, 21810--21825}~(2022).

\bibitem{Intro20212}
A.~Laucht, F.~Hohls, N.~Ubbelohde, M.~F. Gonzalez-Zalba, D.~J. Reilly, S.~Stobbe, T.~Schröder, P.~Scarlino, J.~V. Koski, A.~Dzurak, C.-H. Yang, J.~Yoneda, F.~Kuemmeth, H.~Bluhm, J.~Pla, C.~Hill, J.~Salfi, A.~Oiwa, J.~T. Muhonen, E.~Verhagen, M.~D. LaHaye, H.~H. Kim, A.~W. Tsen, D.~Culcer, A.~Geresdi, J.~A. Mol, V.~Mohan, P.~K. Jain, and J.~Baugh.
\newblock ``Roadmap on quantum nanotechnologies''.
\newblock \href{https://dx.doi.org/https://doi.org/10.1088/1361-6528/abb333}{Nanotechnology {\bf 32}, 162003}~(2021).

\bibitem{Intro20213}
D.~S. Wang, M.~Haas, and P.~Narang.
\newblock ``Quantum interfaces to the nanoscale''.
\newblock \href{https://dx.doi.org/https://doi.org/10.1021/acsnano.1c01255}{ACS Nano {\bf 15}, 7879–7888}~(2021).

\bibitem{Intro20172}
L.~Schlipf, T.~Oeckinghaus, K.~Xu, D.~B.~R. Dasari, A.~Zappe, F.~F. de~Oliveira, B.~Kern, M.~Azarkh, M.~Drescher, M.~Ternes, K.~Kern, J.~Warchtrup, and A.~Finkler.
\newblock ``A molecular quantum spin network controlled by a single qubit''.
\newblock \href{https://dx.doi.org/https://doi.org/10.1126/sciadv.1701116}{Sci. Adv.{\bf 3}}~(2017).

\bibitem{Intro2018}
D.~D. Awschalom, R.~Hanson, J.~Wrachtrup, and B.~B. Zhou.
\newblock ``Quantum technologies with optically interfaced solid-state spins''.
\newblock \href{https://dx.doi.org/https://doi.org/10.1038/s41566-018-0232-2}{Nat. Photonics {\bf 12}, 516–527}~(2018).

\bibitem{Intro2022}
M.~T. Madzik, S.~Asaad, A.~Youssry, B.~Joecker, K.~M. Rudinger, E.~Nielsen, K.~C. Young, T.~J. Proctor, A.~D. Baczewski, A.~Laucht, V.~Schmitt, F.~E. Hudson, K.~M. Itoh, A.~M. Jakob, B.~C. Johnson, D.~N. Jamieson, A.~S. Dzurak, C.~Ferrie, R.~Blume-Kohout, and A.~Morello.
\newblock ``Precision tomography of a three-qubit donor quantum processor in silicon''.
\newblock \href{https://dx.doi.org/https://doi.org/10.1038/s41586-021-04292-7}{Nature {\bf 601}, 348–353}~(2022).

\bibitem{Review2019}
M.~Atzori and R.~Sessoli.
\newblock ``The second quantum revolution: Role and challenges of molecular chemistry''.
\newblock \href{https://dx.doi.org/https://doi.org/10.1021/jacs.9b00984}{J. Am. Chem. Soc. {\bf 141}, 11339–11352}~(2019).

\bibitem{Review20201}
E.~Coronado.
\newblock ``Molecular magnetism: from chemical design to spin control in molecules, materials and devices''.
\newblock \href{https://dx.doi.org/https://doi.org/10.1038/s41578-019-0146-8}{Nat. Rev. Mater. {\bf 5}, 87–104}~(2020).

\bibitem{Review20202}
M.~R. Wasielewski, M.~D.~E. Forbes, N.~L. Frank, K.~Kowalski, G.~D. Scholes, J.~Yuen-Zhou, M.~A. Baldo, D.~E. Freedman, R.~H. Goldsmith, T.~Goodson III, M.~L. Kirk, J.~K. McCusker, J.~P. Ogilvie, D.~A. Shultz, S.~Stoll, and K.~B. Whaley.
\newblock ``Exploiting chemistry and molecular systems for quantum information science''.
\newblock \href{https://dx.doi.org/https://doi.org/10.1038/s41570-020-0200-5}{Nat. Rev. Chem. {\bf 4}, 490–504}~(2020).

\bibitem{Imaging20211}
P.~Willke, T.~Bilgeri, X.~Zhang, Y.~Wang, C.~Wolf, H.~Aubin, A.~Heinrich, and T.~Choi.
\newblock ``Coherent spin control of single molecules on a surface''.
\newblock \href{https://dx.doi.org/https://doi.org/10.1021/acsnano.1c06394}{ACS Nano {\bf 15}, 17959–17965}~(2021).

\bibitem{Imaging20213}
S.~Lenz, D.~König, D.~Hunger, and J.~van Slageren.
\newblock ``Room-temperature quantum memories based on molecular electron spin ensembles''.
\newblock \href{https://dx.doi.org/https://doi.org/10.1002/adma.202101673}{Adv. Mater. {\bf 33}, 2101673}~(2021).

\bibitem{Qudits20211}
I.~Gimeno, A.~Urtizberea, J.~Román-Roche, D.~Zueco, A.~Camón, P.~J. Alonso, O.~Roubeau, and F.~Luis.
\newblock ``Broad-band spectroscopy of a vanadyl porphyrin: a model electronuclear spin qudit''.
\newblock \href{https://dx.doi.org/https://doi.org/10.1039/D1SC00564B}{Chem. Sci. {\bf 12}, 5621--5630}~(2021).

\bibitem{Qudits20213}
H.~Biard, E.~Moreno-Pineda, M.~Ruben, E.~Bonet, W.~Wernsdorfer, and F.~Balestro.
\newblock ``Increasing the hilbert space dimension using a single coupled molecular spin''.
\newblock \href{https://dx.doi.org/https://doi.org/10.1038/s41467-021-24693-6}{Nat. Commun. {\bf 12}, 4443}~(2021).

\bibitem{Optimal20221}
T.~Shibata, S.~Yamamoto, S.~Nakazawa, E.~H. Lapasar, K.~Sugisaki, K.~Maruyama, K.~Toyota, D.~Shiomi, K.~Sato, and T.~Takui.
\newblock ``Molecular optimization for nuclear spin state control via a single electron spin qubit by optimal microwave pulses: Quantum control of molecular spin qubits''.
\newblock \href{https://dx.doi.org/https://doi.org/10.1007/s00723-021-01392-5}{Appl. Magn. Reson. {\bf 53}, 777–796}~(2022).

\bibitem{Optimal20222}
A.~Castro, A.~G. Carrizo, S.~Roca, D.~Zueco, and F.~Luis.
\newblock ``Optimal control of molecular spin qudits''.
\newblock \href{https://dx.doi.org/https://doi.org/10.1103/PhysRevApplied.17.064028}{Phys. Rev. Applied {\bf 17}, 064028}~(2022).

\bibitem{Puertas20222}
A.~Ullah, Z.~Hu, J.~Cerdà, J.~Aragó, and A.~Gaita-Ariño.
\newblock ``Electrical two-qubit gates within a pair of clock-qubit magnetic molecules''.
\newblock \href{https://dx.doi.org/https://doi.org/10.1038/s41534-022-00647-8}{npj Quantum Inf. {\bf 8}, 133}~(2022).

\bibitem{Imaging20172}
E.~Garlatti, T.~Guidi, S.~Ansbro, P.~Santini, G.~Amoretti, J.~Ollivier, H.~Mutka, G.~Timco, I.~J. Vitorica-Yrezabal, G.~F.~S. Whitehead, R.~E.~P. Winpenny, and S.~Carretta.
\newblock ``Portraying entanglement between molecular qubits with four-dimensional inelastic neutron scattering''.
\newblock \href{https://dx.doi.org/https://doi.org/10.1038/ncomms14543}{Nat. Commun. {\bf 8}, 14543}~(2021).

\bibitem{Puertas2017}
C.~Godfrin, A.~Ferhat, R.~Ballou, S.~Klyatskaya, M.~Ruben, W.~Wernsdorfer, and F.~Balestro.
\newblock ``Operating quantum states in single magnetic molecules: Implementation of \uppercase{G}rover’s quantum algorithm''.
\newblock \href{https://dx.doi.org/https://doi.org/10.1103/PhysRevLett.119.187702}{Phys. Rev. Lett. {\bf 119}, 187702}~(2017).

\bibitem{Escalado2021}
S.~Carretta, D.~Zueco, A.~Chiesa, A.~Gómez-León, and F.~Luis.
\newblock ``A perspective on scaling up quantum computation with molecular spins''.
\newblock \href{https://dx.doi.org/https://doi.org/10.1063/5.0053378}{Appl. Phys. Lett. {\bf 118}, 240501}~(2021).

\bibitem{QEC20222}
A.~Chiesa, F.~Petiziol, M.~Chizzini, P.~Santini, and S.~Carretta.
\newblock ``Theoretical design of optimal molecular qudits for quantum error correction''.
\newblock \href{https://dx.doi.org/https://doi.org/10.1021/acs.jpclett.2c01602}{J. Phys. Chem. Lett. {\bf 13}, 6468--6474}~(2022).

\bibitem{Overview2024}
S.~Chicco, G.~Allodi, A.~Chiesa, E.~Garlatti, C.~D. Buch, P.~Santini, R.~De Renzi, S.~Piligkos, and S.~Carretta.
\newblock ``Proof-of-concept quantum simulator based on molecular spin qudits''.
\newblock \href{https://dx.doi.org/https://doi.org/10.1021/jacs.3c12008}{J. Am. Chem. Soc. {\bf 146}, 1053--1061}~(2024).

\bibitem{Overview2020}
D.~Aravena and E.~Ruiz.
\newblock ``Spin dynamics in single-molecule magnets and molecular qubits''.
\newblock \href{https://dx.doi.org/https://doi.org/10.1039/D0DT01414A}{Dalton Trans. {\bf 49}, 9916--9928}~(2020).

\bibitem{Overview2021}
R.~Mirzoyan, N.~P. Kazmierczak, and R.~G. Hadt.
\newblock ``Deconvolving contributions to decoherence in molecular electron spin qubits: A dynamic ligand field approach''.
\newblock \href{https://dx.doi.org/https://doi.org/10.1002/chem.202100845}{Chem. Eur. J. {\bf 27}, 9482--9494}~(2021).

\bibitem{Overview2022}
A.~Lunghi and S.~Sanvito.
\newblock ``Computational design of magnetic molecules and their environment using quantum chemistry, machine learning and multiscale simulations''.
\newblock \href{https://dx.doi.org/https://doi.org/10.1038/s41570-022-00424-3}{Nat. Rev. Chem. {\bf 6}, 761–781}~(2022).

\bibitem{espesp2016}
S.~Cardona-Serra, L.~Escalera-Moreno, J.~J. Baldoví, A.~Gaita-Ariño, J.~M. Clemente-Juan, and E.~Coronado.
\newblock ``\uppercase{SIMPRE}1.2: Considering the hyperfine and quadrupolar couplings and the nuclear spin bath decoherence''.
\newblock \href{https://dx.doi.org/https://doi.org/10.1002/jcc.24313}{J. Comput. Chem. {\bf 37}, 1238--1244}~(2016).

\bibitem{Maestra2017}
S.~Lenz, K.~Bader, H.~Bamberger, and J.~van Slageren.
\newblock ``Quantitative prediction of nuclear-spin-diffusion-limited coherence times of molecular quantum bits based on copper(\uppercase{II})''.
\newblock \href{https://dx.doi.org/https://doi.org/10.1039/C6CC07813C}{Chem. Commun. {\bf 53}, 4477}~(2017).

\bibitem{espesp2019}
L.~Escalera-Moreno, A.~Gaita-Ariño, and E.~Coronado.
\newblock ``Decoherence from dipolar interspin interactions in molecular spin qubits''.
\newblock \href{https://dx.doi.org/https://doi.org/10.1103/PhysRevB.100.064405}{Phys. Rev. B {\bf 100}, 064405}~(2019).

\bibitem{Maestra20191}
A.~Lunghi and S.~Sanvito.
\newblock ``Electronic spin-spin decoherence contribution in molecular qubits by quantum unitary spin dynamics''.
\newblock \href{https://dx.doi.org/https://doi.org/10.1016/j.jmmm.2019.165325}{J. Magn. Magn. Mater. {\bf 487}, 165325}~(2019).

\bibitem{espesp20201}
J.~Chen, C.~Hu, J.~F. Stanton, S.~Hill, H.-Ping Cheng, and X.-Guang Zhang.
\newblock ``Decoherence in molecular electron spin qubits: Insights from quantum many-body simulations''.
\newblock \href{https://dx.doi.org/https://doi.org/10.1021/acs.jpclett.0c00193}{J. Phys. Chem. Lett. {\bf 11}, 2074–2078}~(2020).

\bibitem{espesp20202}
E.~R. Canarie, S.~M. Jahn, and S.~Stoll.
\newblock ``Quantitative structure-based prediction of electron spin decoherence in organic radicals''.
\newblock \href{https://dx.doi.org/https://doi.org/10.1021/acs.jpclett.0c00768}{J. Phys. Chem. Lett. {\bf 11}, 3396–3400}~(2020).

\bibitem{espesp20221}
S.~Kanai, F.~J. Heremans, H.~Seo, and H.~Ohno.
\newblock ``Generalized scaling of spin qubit coherence in over 12,000 host materials''.
\newblock \href{https://dx.doi.org/https://doi.org/10.1073/pnas.2121808119}{Proc. Natl. Acad. Sci. U.S.A {\bf 119}, 15}~(2022).

\bibitem{espesp20222}
X.-Fa Jiang, Z.-Bo Hu, C.~Shao, Z.~Ouyang, Z.~Wang, and Y.~Song.
\newblock ``Molecular spin qubits impregnated in a hexagonal self-ordered mesoporous silica''.
\newblock \href{https://dx.doi.org/https://doi.org/10.1021/acs.chemmater.2c02181}{Chem. Mater. {\bf 34}, 8427–8436}~(2022).

\bibitem{espesp2023}
K.~Kundu, J.~Chen, S.~Hoffman, J.~Marbey, D.~Komijani, Y.~Duan, A.~Gaita-Ariño, J.~Stanton, X.~Zhang, H.-Ping Cheng, and S.~Hill.
\newblock ``Electron-nuclear decoupling at a spin clock transition''.
\newblock \href{https://dx.doi.org/https://doi.org/10.1038/s42005-023-01152-w}{Commun. Phys. {\bf 6}, 38}~(2023).

\bibitem{espvib20173}
A.~Lunghi, F.~Totti, R.~Sessoli, and S.~Sanvito.
\newblock ``The role of anharmonic phonons in under-barrier spin relaxation of single molecule magnets''.
\newblock \href{https://dx.doi.org/https://doi.org/10.1038/ncomms14620}{Nat. Commun. {\bf 8}, 14620}~(2017).

\bibitem{espvib2018}
L.~Escalera-Moreno, J.~J. Baldoví, A.~Gaita-Ariño, and E.~Coronado.
\newblock ``Spin states, vibrations and spin relaxation in molecular nanomagnets and spin qubits: a critical perspective''.
\newblock \href{https://dx.doi.org/https://doi.org/10.1039/C7SC05464E}{Chem. Sci. {\bf 9}, 3265--3275}~(2018).

\bibitem{espvib2019}
A.~Lunghi and S.~Sanvito.
\newblock ``How do phonons relax molecular spins?''.
\newblock \href{https://dx.doi.org/https://doi.org/10.1126/sciadv.aax7163}{Sci. Adv. {\bf 5}, 9}~(2019).

\bibitem{espvib20211}
M.~Briganti, F.~Santanni, L.~Tesi, F.~Totti, R.~Sessoli, and A.~Lunghi.
\newblock ``A complete \uppercase{A}b \uppercase{I}nitio view of \uppercase{O}rbach and \uppercase{R}aman spin–lattice relaxation in a dysprosium coordination compound''.
\newblock \href{https://dx.doi.org/https://doi.org/10.1021/jacs.1c05068}{J. Am. Chem. Soc. {\bf 143}, 13633–13645}~(2021).

\bibitem{espvib20212}
D.~Reta, J.~G.~C. Kragskow, and N.~F. Chilton.
\newblock ``Ab \uppercase{I}nitio prediction of high-temperature magnetic relaxation rates in single-molecule magnets''.
\newblock \href{https://dx.doi.org/https://doi.org/10.1021/jacs.1c01410}{J. Am. Chem. Soc. {\bf 143}, 5943–5950}~(2021).

\bibitem{espvib20223}
A.~Lunghi.
\newblock ``Toward exact predictions of spin-phonon relaxation times: An ab initio implementation of open quantum systems theory''.
\newblock \href{https://dx.doi.org/https://doi.org/10.1126/sciadv.abn7880}{Sci. Adv. {\bf 8}, 31}~(2022).

\bibitem{espvib20224}
S.~Mondal and A.~Lunghi.
\newblock ``Unraveling the contributions to spin–lattice relaxation in \uppercase{K}ramers single-molecule magnets''.
\newblock \href{https://dx.doi.org/https://doi.org/10.1021/jacs.2c08876}{J. Am. Chem. Soc. {\bf 144}, 22965–22975}~(2022).

\bibitem{espvib2023}
R.~Nabi, J.~K. Staab, A.~Mattioni, J.~G.~C. Kragskow, D.~Reta, J.~M. Skelton, and N.~F. Chilton.
\newblock ``Accurate and efficient spin–phonon coupling and spin dynamics calculations for molecular solids''.
\newblock \href{https://dx.doi.org/https://doi.org/10.1021/jacs.3c06015}{J. Am. Chem. Soc. {\bf 145}, 24558--24567}~(2023).

\bibitem{espvib20171}
L.~Escalera-Moreno, N.~Suaud, A.~Gaita-Ariño, and E.~Coronado.
\newblock ``Determining key local vibrations in the relaxation of molecular spin qubits and single-molecule magnets''.
\newblock \href{https://dx.doi.org/https://doi.org/10.1021/acs.jpclett.7b00479}{J. Phys. Chem. Lett. {\bf 8}, 1695–1700}~(2017).

\bibitem{espvib20172}
A.~Lunghi, F.~Totti, S.~Sanvito, and R.~Sessoli.
\newblock ``Intra-molecular origin of the spin-phonon coupling in slow-relaxing molecular magnets''.
\newblock \href{https://dx.doi.org/https://doi.org/10.1039/C7SC02832F}{Chem. Sci. {\bf 8}, 6051--6059}~(2017).

\bibitem{espvib2020}
A.~Lunghi and S.~Sanvito.
\newblock ``The limit of spin lifetime in solid-state electronic spins''.
\newblock \href{https://dx.doi.org/https://doi.org/10.1021/acs.jpclett.0c01681}{J. Phys. Chem. Lett. {\bf 11}, 6273–6278}~(2020).

\bibitem{espvib20221}
V.~H.~A. Nguyen and A.~Lunghi.
\newblock ``Predicting tensorial molecular properties with equivariant machine learning models''.
\newblock \href{https://dx.doi.org/https://doi.org/10.1103/PhysRevB.105.165131}{Phys. Rev. B {\bf 105}, 165131}~(2022).

\bibitem{Maestra20194}
A.~Albino, S.~Benci, L.~Tesi, M.~Atzori, R.~Torre, S.~Sanvito, R.~Sessoli, and A.~Lunghi.
\newblock ``First-principles investigation of spin–phonon coupling in vanadium-based molecular spin quantum bits''.
\newblock \href{https://dx.doi.org/https://doi.org/10.1021/acs.inorgchem.9b01407}{Inorg. Chem. {\bf 58}, 10260–10268}~(2019).

\bibitem{Sistemas20162}
M.~Atzori, E.~Morra, L.~Tesi, A.~Albino, M.~Chiesa, L.~Sorace, and R.~Sessoli.
\newblock ``Quantum coherence times enhancement in vanadium(\uppercase{IV})-based potential molecular qubits: the key role of the vanadyl moiety''.
\newblock \href{https://dx.doi.org/https://doi.org/10.1021/jacs.6b05574}{J. Am. Chem. Soc. {\bf 138}, 11234–11244}~(2016).

\bibitem{Sistemas20161}
M.~Atzori, L.~Tesi, E.~Morra, M.~Chiesa, L.~Sorace, and R.~Sessoli.
\newblock ``Room-temperature quantum coherence and \uppercase{R}abi oscillations in vanadyl phthalocyanine: Toward multifunctional molecular spin qubits''.
\newblock \href{https://dx.doi.org/https://doi.org/10.1021/jacs.5b13408}{J. Am. Chem. Soc. {\bf 138}, 2154–2157}~(2016).

\bibitem{Sistemas2014}
K.~Bader, D.~Dengler, S.~Lenz, B.~Endeward, S.-Da Jiang, P.~Neugebauer, and J.~van Slageren.
\newblock ``Room temperature quantum coherence in a potential molecular qubit''.
\newblock \href{https://dx.doi.org/https://doi.org/10.1038/ncomms6304}{Nat. Commun. {\bf 5}, 5304}~(2014).

\bibitem{Sistemas2017}
Y.~Dai, Z.~Shi, Y.~Fu, X.~Qin, S.~Mu, Y.~Wu, J.-Hu Su, L.~Qin, Y.-Qi Zhai, Y.-Fei Deng, X.~Rong, and J.~Du.
\newblock ``Experimental protection of the coherence of a molecular qubit exceeding a millisecond \href{https://doi.org/10.48550/arXiv.1706.09259}{arXiv:1706.09259}''~(2017).

\bibitem{Maestra2018}
A.~Norambuena, E.~Muñoz, H.~T. Dinani, A.~Jarmola, P.~Maletinsky, D.~Budker, and J.~R. Maze.
\newblock ``Spin-lattice relaxation of individual solid-state spins''.
\newblock \href{https://dx.doi.org/https://doi.org/10.1103/PhysRevB.97.094304}{Phys. Rev. B {\bf 97}, 094304}~(2018).

\bibitem{Maestra2015}
J.~E. Avron, O.~Kenneth, A.~Retzker, and M.~Shalyt.
\newblock ``Lindbladians for controlled stochastic hamiltonians''.
\newblock \href{https://dx.doi.org/https://doi.org/10.1088/1367-2630/17/4/043009}{New J. Phys. {\bf 17}, 043009}~(2015).

\bibitem{millicohtime}
J.~M. Zadrozny, J.~Niklas, O.~G. Poluektov, and D.~E. Freedman.
\newblock ``Millisecond coherence time in a tunable molecular electronic spin qubit''.
\newblock \href{https://dx.doi.org/https://doi.org/10.1021/acscentsci.5b00338}{ACS Cent. Sci. {\bf 1}, 488--492}~(2015).

\bibitem{Maestra20222}
M.~J. Amdur, K.~R. Mullin, M.~J. Waters, D.~Puggioni, M.~K. Wojnar, M.~Gu, L.~Sun, P.~H. Oyala, J.~M. Rondinelli, and D.~E. Freedman.
\newblock ``Chemical control of spin–lattice relaxation to discover a room temperature molecular qubit''.
\newblock \href{https://dx.doi.org/https://doi.org/10.1039/D1SC06130E}{Chem. Sci. {\bf 13}, 7034--7045}~(2022).

\bibitem{Maestra20221}
L.~Gu, J.~Li, and R.~Wu.
\newblock ``Reconsidering spin-phonon relaxation in magnetic molecules''.
\newblock \href{https://dx.doi.org/https://doi.org/10.1016/j.jmmm.2022.170138}{J. Magn. Magn. Mater. {\bf 564}, 170138}~(2022).

\bibitem{gapwidth2016}
M.~Shiddiq, D.~Komijani, Y.~Duan, A.~Gaita-Ariño, E.~Coronado, and S.~Hill.
\newblock ``Enhancing coherence in molecular spin qubits via atomic clock transitions''.
\newblock \href{https://dx.doi.org/https://doi.org/10.1038/nature16984}{Nature {\bf 531}, 348–351}~(2016).

\bibitem{ElecCont20211}
J.~Liu, J.~Mrozek, A.~Ullah, Y.~Duan, J.~J. Baldoví, E.~Coronado, A.~Gaita-Ariño, and A.~Ardavan.
\newblock ``Quantum coherent spin–electric control in a molecular nanomagnet at clock transitions''.
\newblock \href{https://dx.doi.org/https://doi.org/10.1038/s41567-021-01355-4}{Nat. Phys. {\bf 17}, 1205–1209}~(2021).

\bibitem{Fidel2022}
T.~Abad, A.~F. Kockum, and G.~Johansson.
\newblock ``Impact of decoherence on the fidelity of quantum gates leaving the computational subspace \href{https://doi.org/10.48550/arXiv.2302.13885}{arXiv:2302.13885}''~(2024).

\end{thebibliography}

\end{document}